\documentclass[a4paper,twoside,11pt,aps,superscriptaddress,amsmath,amsfonts,amssymb,pre,floatfix,longbibliography,nofootinbib]{revtex4-1}

\pdfoutput=1

\usepackage{amsmath}
\usepackage{amssymb}
\usepackage{bbold} 

\usepackage{color}
\usepackage[dvipsnames]{xcolor}
\definecolor{nblue}{RGB}{28,130,185}
\definecolor{cgreen}{RGB}{76,153,0}
\definecolor{bostonuniversityred}{rgb}{0.8, 0.0, 0.0}
\definecolor{myorange}{RGB}{245,156,74}

\usepackage{hyperref}
\hypersetup{
  colorlinks=true,
  citecolor=magenta,
  urlcolor=-myorange
}

\usepackage[utf8]{inputenc}
\usepackage{subfigure}
\usepackage{xifthen}
\usepackage{graphicx}
\usepackage{hyphenat}
\usepackage{hhline}
\usepackage{upgreek}
\usepackage{bm}
\usepackage{xcolor}

\usepackage{subfiles} 

\newcommand{\alabel}[1]{\label{app:#1}}

\newcommand{\rmlabel}[3]{\ensuremath{#1_\mathrm{#2}^\mathrm{#3}}}

\newcommand{\nematicT}[2]{%
\ifthenelse{\isempty{#2}}{\ensuremath{\widetilde{#1}_{ij}}}{\ensuremath{\widetilde{#1}_{#2}}}%
}
\newcommand{\normalT}[2]{%
\ifthenelse{\isempty{#2}}{\ensuremath{#1_{ij}}}{\ensuremath{#1_{#2}}}%
}

\newcommand{\shearmod}[0]{\ensuremath{\rmlabel{\mu}{s}{}}}

\newcommand{\shearv}[1]{%
\ifthenelse{\isempty{#1}}{\nematicT{v}{}}{\nematicT{v}{#1}}%
}
\newcommand{\strainv}[1]{%
\ifthenelse{\isempty{#1}}{\normalT{v}{}}{\normalT{v}{#1}}%
}

\newcommand{\shears}[1]{%
\ifthenelse{\isempty{#1}}{\nematicT{\sigma}{}}{\nematicT{\sigma}{#1}}%
}
\newcommand{\shearsss}[1]{%
\ifthenelse{\isempty{#1}}{\nematicT{\sigma}{}^\mathrm{ss}}{\nematicT{\sigma}{#1}^\mathrm{ss}}%
}
\newcommand{\shearas}[1]{%
\ifthenelse{\isempty{#1}}{\normalT{\tilde\sigma^{\mathrm{a}}}{}}{\normalT{\tilde\sigma^{\mathrm{a}}}{#1}}%
}
\newcommand{\shearQVM}[1]{%
\ifthenelse{\isempty{#1}}{\ensuremath{\normalT{G}{}}}{\ensuremath{\normalT{G}{#1}}}%
}
\newcommand{\shearQ}[1]{%
\ifthenelse{\isempty{#1}}{\normalT{\widetilde{Q}}{}}{\normalT{\widetilde{Q}}{#1}}%
}
\newcommand{\shearQss}[1]{%
\ifthenelse{\isempty{#1}}{\ensuremath{\normalT{\widetilde{Q}}{}^{\mathrm{ss}}}}{\ensuremath{\normalT{\widetilde{Q}}{#1}^{\mathrm{ss}}}}%
}
\newcommand{\shearq}[1]{%
\ifthenelse{\isempty{#1}}{\normalT{\widetilde{q}}{}}{\normalT{\widetilde{q}}{#1}}%
}
\newcommand{\shearR}[1]{%
\ifthenelse{\isempty{#1}}{\normalT{R}{}}{\normalT{R}{#1}}%
}
\newcommand{\shearp}[1]{%
\ifthenelse{\isempty{#1}}{\normalT{\widetilde{p}}{}}{\normalT{\widetilde{p}}{#1}}%
}
\newcommand{\shearDQDt}[1]{%
\ifthenelse{\isempty{#1}}{\ensuremath{\frac{D\widetilde{Q}_{ij}}{Dt}}}{\ensuremath{\frac{D\widetilde{Q}_{#1}}{Dt}}}%
}
\newcommand{\stress}[1]{%
\ifthenelse{\isempty{#1}}{\normalT{\sigma}{}}{\normalT{\sigma}{#1}}%
}

\newcommand{\DDt}[1]{%
\frac{{\rm D}#1}{{\rm D}t}%
}
\newcommand{\DDtinline}[1]{%
{\rm D}#1/{\rm D}t%
}

\newcommand{\ddt}[1]{%
\frac{{\rm d}#1}{{\rm d}t}%
}
\newcommand{\ddtinline}[1]{%
{\rm d}#1/{\rm d}t%
}

\begin{document}

\title{Nonlinear rheology of cellular networks}

\author{Charlie~\surname{Duclut}}
\thanks{These authors contributed equally.}
\affiliation{Max Planck Institute for the Physics of Complex Systems, N\"othnitzer Str. 8, 01187 Dresden, Germany}
\author{Joris~\surname{Paijmans}}
\thanks{These authors contributed equally.}
\affiliation{Max Planck Institute for the Physics of Complex Systems, N\"othnitzer Str. 8, 01187 Dresden, Germany}
\author{Mandar~M.~\surname{Inamdar}}
\affiliation{Department of Civil Engineering, Indian Institute of Technology Bombay, Powai, Mumbai 400076, India}
\author{Carl~D.~\surname{Modes}}
\affiliation{Max Planck Institute for Molecular Cell Biology and Genetics (MPI-CBG), Dresden 01307, Germany}
\affiliation{Center for Systems Biology Dresden, Pfotenhauerstrasse 108, 01307 Dresden, Germany}
\affiliation{Cluster of Excellence, Physics of Life, TU Dresden, Dresden 01307, Germany}
\author{Frank~\surname{J\"ulicher}}
\affiliation{Max Planck Institute for the Physics of Complex Systems, N\"othnitzer Str. 8, 01187 Dresden, Germany}
\affiliation{Center for Systems Biology Dresden, Pfotenhauerstrasse 108, 01307 Dresden, Germany}
\affiliation{Cluster of Excellence, Physics of Life, TU Dresden, Dresden 01307, Germany}

\begin{abstract}
Morphogenesis depends crucially on the complex rheological properties of cell tissues and on their ability to maintain mechanical integrity while rearranging at long times.
In this paper,
we study the rheology of polygonal cellular networks described by a vertex model in the presence of fluctuations. 
We use a triangulation method to decompose shear into cell shape changes and cell rearrangements. 
%
Considering the steady-state stress under constant shear, we observe nonlinear shear-thinning behavior at all magnitudes of the fluctuations, and an even stronger nonlinear regime at lower values of the fluctuations. We successfully capture this nonlinear rheology by a mean-field model that describes the tissue in terms of cell elongation and cell rearrangements. We furthermore introduce anisotropic active stresses in the vertex model and analyze their effect on rheology.
We include this anisotropy in the mean-field model and show that it recapitulates the behavior observed in the simulations.
Our work clarifies how tissue rheology is related to stochastic cell rearrangements and provides a simple biophysical model to describe biological tissues. Further, it highlights the importance of nonlinearities when discussing tissue mechanics.
\end{abstract}

\maketitle


\section{Introduction}

During morphogenesis, cells divide, die, rearrange, and flow to create complex structures and shape organs. On short time scales, cells maintain tissue mechanical integrity and form a solid-like structure, while at longer time scales, tissues can deform and relax internal stresses, thus behaving as viscous fluids.
This long time-scale fluidization in confluent tissues occurs as a consequence of topological rearrangements of the cellular network  such as cell divisions, cell extrusions and cell neighbor exchanges (T1 transitions)~\cite{guirao2015unified,etournay2015interplay}.
The understanding of the rheological properties of tissues and their viscoelastic behavior is therefore crucial to link short-time and cell-based events to large-scale and long-time morphogenesis.

In confluent tissues, fluctuations play a significant role in tissue fluidization by facilitating the relaxation of local stresses~\cite{beysens2000cell,marmottant2009role,curran2017myosin,tetley2019tissue}. Tension fluctuations, which have for instance been studied in detail in \textit{Drosophila}~\cite{curran2017myosin} and in zebrafish embryos~\cite{mongera2018fluidtosolid,kim2020embryonic}, are especially relevant as they can trigger T1 transitions and therefore lead to tissue flows. 
The importance of fluctuations for tissue fluidization has also been investigated numerically both with vertex model simulations~\cite{curran2017myosin,bi2016motilitydriven,krajnc2018fluidization,kim2020embryonic,sussman2018anomalous,yamamoto2020nonmonotonic} and cellular Potts model simulations~\cite{marmottant2009role,chiang2016glass}.
These fluctuations, introduced as a noise in the vertex positions~\cite{sussman2018anomalous} or in the form of bond tension fluctuations~\cite{curran2017myosin,krajnc2018fluidization,yamamoto2020nonmonotonic,kim2020embryonic}, lead in both cases to a glassy transition when the magnitude of the noise is varied.

Great effort has been put into the characterization of these glassy dynamics in terms of state diagrams, and their signature on cell motion has been studied extensively, for instance in terms of mean-square displacement of the particles or effective diffusivity~\cite{chiang2016glass,bi2016motilitydriven,kim2020embryonic}. However, the role of tension fluctuations and active noises on the macroscopic rheological properties of cell networks remains poorly characterized, and a nonlinear coarse-grained description of these systems that would capture the glassy transition is still missing. 
Coarse-grained descriptions of tissues, using for instance tools from active gels and active hydrodynamics~\cite{marchetti2013hydrodynamics,prost2015active}, have proven extremely valuable to characterize tissue fluidization by death and growth~\cite{ranft2010fluidization} or to exhibit the role of electric and hydraulic phenomena in tissues~\cite{sarkar2019field,duclut2019fluid,duclut2020hydraulic}. These models can then in turn provide new insights to understand morphogenetic events, for instance the role of mecho-sensitive feedback in the \textit{Drosophila} wing imaginal disc development~\cite{dye2020selforganized}.  

In this theoretical paper, we perform a systematic study of the role of bond tension fluctuations on the rheological properties of cellular networks.
Our analysis is guided by a two-dimensional vertex model, which has been shown to provide a remarkable agreement with experimental data in the case of \textit{Drosophila} wing morphogenesis~\cite{etournay2015interplay} and for other confluent monolayers~\cite{alt2017vertex,fletcher2014vertex,tetley2019tissue,comelles2021epithelial}. 
To quantify the dynamical deformations of the cell network and relate them to cellular processes, we use a shear decomposition based on a triangulation of the cell network~\cite{merkel2017triangles}.
This analysis reveals how large scale tissue shape changes are dominated at short time by cell elongation and at long time by cell rearrangements. We use different triangulations and show that some measures are robust and independent of the triangulation used, while others depend on the choice of triangulation.

Using this approach, we study the relationship between stress and strain rate in tissues.
We observe nonlinearities in the stress versus strain rate relationship of the vertex model for all magnitudes of the fluctuations.
Finally, we show that a generic nonlinear description of the material properties of cell networks can account for them.

The paper is organized as follows. In Sec.~\ref{sec_vertex_model}, we introduce the vertex model and its work function. Using two different triangulation methods, we then analyze the dynamics of the system under pure shear in Sec.~\ref{sec_dynamics}. We show that the dynamics has nonlinear effects that cannot be captured by a linear continuum description. We then introduce and discuss a nonlinear model in Sec.~\ref{sec_non_linear_response}, both in the case of an isotropic cell network and in the case of a network with an intrinsic anisotropy.

\section{Mechanics of polygonal cell networks}

\label{sec_vertex_model}

The cell boundaries, defined as the network of apical cellular junctions of epithelial tissues, form a 2D network that is well-represented by a packing of convex polygons. The mechanics of this network can be described by a vertex model, where cells are represented as polygons that are outlined by straight edges connecting vertices~\cite{farhadifar2007influence}. In this section, we briefly review the vertex model and describe the dynamics of the polygonal cell network arising from bond tension fluctuations, topological rearrangements in the network, and dynamic boundary conditions.

\subsection{Work function of the cell network}

\begin{figure}[t]
 \centering
 \includegraphics[width=1.0\textwidth]{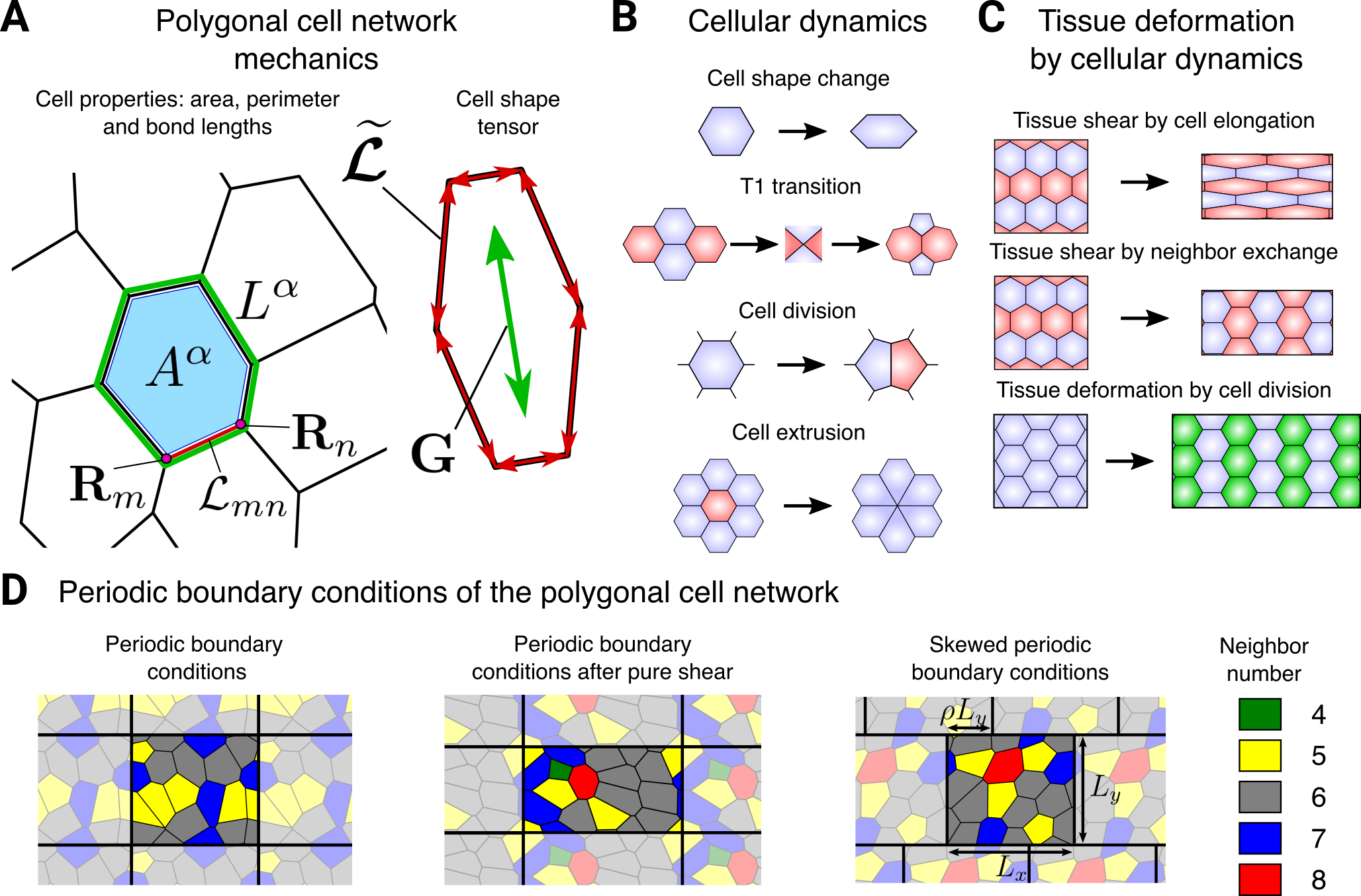}
 \caption{
Mechanics and dynamics of cellular networks. 
\textbf{(A)} Definition of the cell state variables. 
Left shows the cell area $A^\alpha$ (blue patch), cell perimeter $L^\alpha$ (green line) and bond length $\mathcal{L}_{mn}$ (red line) between the vertices with positions $\mathbf{R}_m$ and $\mathbf{R}_{n}$.
Right shows the the cell elongation tensor $\mathbf{G}$ which is constructed from the bond nematic tensors $\bm{\widetilde{\mathcal{L}}}$ as defined in Eq.~\eqref{cell_shape_tensor}. 
\textbf{(B)} Cell dynamic processes can lead to tissue deformation as an effect of cell shape changes, T1 transitions, cell divisions or cell extrusions.
\textbf{(C)} Large scale tissue deformation can be driven by collective cell dynamics: 
cell shape changes (top), anisotropic T1 transitions (middle) and anisotropic cell divisions (bottom). The tissue may deform as a result of changes in the mean cell shape of the cellular network. Alternatively, when T1 transitions preferably occur on edges with a certain orientation, the resulting rearrangement of the cells can induce tissue shear.
Cell divisions with a preferred division axis (vertical in the figure), the division and the subsequent growth of the cells can induce a tissue deformation (new cells are colored green).
\textbf{(D)} Boundary conditions used in this paper. Cell colors indicate the neighbor number, given in the legend to the right. Adapted from Ref.~\cite{merkel2014cells}.
 }
 \label{fig_polygonal_cell_network}
\end{figure}

We consider a polygonal cell network consisting of \rmlabel{N}{c}{} polygonal cells labeled $\alpha$, \rmlabel{N}{v}{} vertices labeled $m$ and \rmlabel{N}{b}{} straight bonds between connected vertices labeled $\langle mn \rangle$ where $m$ and $n$ are the vertices they connect. Each cell is characterized in terms of its area $A^\alpha$, its perimeter $L^\alpha$ and the lengths of the bonds that form the outline of the cell $\mathcal{L}_{mn}$, as shown in Fig.~\ref{fig_polygonal_cell_network}.

We employ a quasistatic representation of epithelia where the cell network is at any instant in a mechanical equilibrium, while the parameters describing cell properties can slowly change with time.
At each vertex $m$, the total force $\mathbf{F}_m = -\partial W_0/\partial \mathbf{R}_m$ vanishes, where $\mathbf{R}_m$ is the position of the vertex, and $W_0$ is the vertex model work function and reads~\cite{honda1984computer, farhadifar2007influence}:
\begin{equation}
  W_0 = \sum_\alpha \frac{1}{2} K^\alpha\left( A^\alpha - A^\alpha_0 \right)^2 + \sum_{\langle m,n \rangle} \Lambda_{mn} \mathcal{L}_{mn} + 
  \sum_\alpha \frac{1}{2} \Gamma^\alpha (L^\alpha)^2.
 \label{eq_tissue_work}
\end{equation}
Note that for clarity, upper-case letters are used here and in the following for quantities related to the vertex model, while lower-case letters will be used for the continuum model.
The first term describes an area elasticity contribution, with $A^\alpha_0$ the preferred cell area and $K^\alpha$ the area stiffness. This term arises from the shear elasticity of cells that keep their volume roughly constant.
The second term describes a contribution due to the tension of network bonds with length $\mathcal{L}_{mn}$ and line tension $\Lambda_{mn}$. This tension is usually positive, favoring the shrinking of the bonds, but could also be negative, favoring an extension of the bonds. The third term describes an elasticity of the cell perimeter with stiffness~$\Gamma^\alpha$.
For simplicity, we use for all cells the same constant values of the parameters $K^\alpha$, $A^\alpha_0$ and $\Gamma^\alpha$. In App.~\ref{sec_vertex_model_implementation}, we give details on the numerical implementation of the vertex model. Values of the (dimensionless) parameters used in the simulations are given in Table~\ref{tab_vertex_model_parameters}.

    \subsection{Bond tension fluctuations and topological network rearrangements}
    
\sloppy

For a fixed connectivity of cells, the cellular network as defined above behaves purely elastically under external deformations. However, tissues are known to undergo plastic deformations and can exhibit viscoelastic behaviors~\cite{guirao2015unified,etournay2015interplay}. This is because they are able to relax internal stresses due to topological rearrangements such as cell neighbor exchanges, cell divisions and cell extrusions \cite{forgacs1998viscoelastic, marmottant2009role, ranft2010fluidization}, see Fig.~\ref{fig_polygonal_cell_network}. 
In the absence of external stresses, topological rearrangements and cell neighbor exchanges (or ``T1 transitions'') can be driven by fluctuations of the bond lengths in the cell network~\cite{curran2017myosin,mongera2018fluidtosolid,tetley2019tissue}. In tissues, these fluctuations could be due to variations in the mechanical properties of cell bonds due to the binding and unbinding of adhesion proteins to the cell membrane or changes of the contractility in the cell cortex. We capture this dynamics by a time-dependent line tension  $\Lambda_{mn}(t)$. 
The line tension dynamics of individual bonds in the network follows an Ornstein-Uhlenbeck process: 
\begin{equation}
 \ddt{\Lambda_{mn}} = -\frac{1}{\tau_\Lambda}(\Lambda_{mn}(t) - \Lambda_0) + \Delta\Lambda \sqrt{2/\tau_\Lambda} \, \Xi_{mn}(t) \, ,
 \label{eq_line_tension_dyn_isotropic}
\end{equation}
where $\Xi_{mn}(t)$ is a Gaussian white noise with zero mean $\langle \Xi_{mn}(t) \rangle = 0$, and correlations \mbox{$\langle \Xi_{mn}(t)\Xi_{op}(t') \rangle = \delta_{\langle mn \rangle,  \langle op \rangle} \delta(t-t')$} 
where $\delta_{\langle mn \rangle,  \langle op \rangle}=1$ if vertices  $\langle mn \rangle $ and $\langle op \rangle$ are the same and 0 otherwise~\cite{farhadifar2007influence, aigouy2010cell}. The line tension of every bond relaxes towards its mean value $\Lambda_0$ with a characteristic time $\tau_\Lambda$, which sets the time scale of the dynamics and is of the order of the acto-myosin cortex turn-over time. Note that this time scale has recently been shown to play a role in the rigidification of the tissue through the formation of ``trapped'' edges~\cite{yamamoto2020nonmonotonic}. 
Finally, the magnitude of bond tension fluctuations $\Delta\Lambda$ has a crucial role in the rheological properties of cell networks, as we discuss in Sec.~\ref{sec_non_linear_response}.

\subsection{Boundary conditions and time dependent shear deformations}

In order to study the rheological properties of the stochastic vertex model, we consider a cell network in a rectangular box with dimensions $L_x$ and $L_y$ with periodic boundary conditions, as shown in Fig.~\ref{fig_polygonal_cell_network}, panel D. 
Throughout the paper, we will apply three different boundary conditions. 
(i)~Under a \textit{fixed} boundary condition, the box dimensions $L_x$ and $L_y$ do not change over time. (ii)~Under a \textit{pure shear} boundary condition, the box dimensions are deformed as \mbox{$L_x= L_x^0\,\exp(\rho)$} and \mbox{$L_y = L_y^0\,\exp(-\rho)$}, such that the aspect-ratio of the box changes but the area remains fixed. At constant shear rate, one has $\dot\rho=V_0$, where the dot stands for the time derivative. (iii)~Under a \textit{simple shear} boundary condition, the dimension of the simulation box are kept fixed and we apply a skewed periodic boundary condition or Lees-Edwards boundary condition~\cite{lees1972computer}: each bond in the network that crosses the horizontal boundary of the periodic box obtains an extra skew $\rho L_y$ in the horizontal direction (see Fig.~\ref{fig_polygonal_cell_network}D). 
Finally, we define the shear stress acting on the simulation box: $\Sigma_0= L_x^{-1}L_y^{-1} \partial W/\partial \rho$.

\begin{figure}[t]
 \centering
 \includegraphics[width=1.0\textwidth]{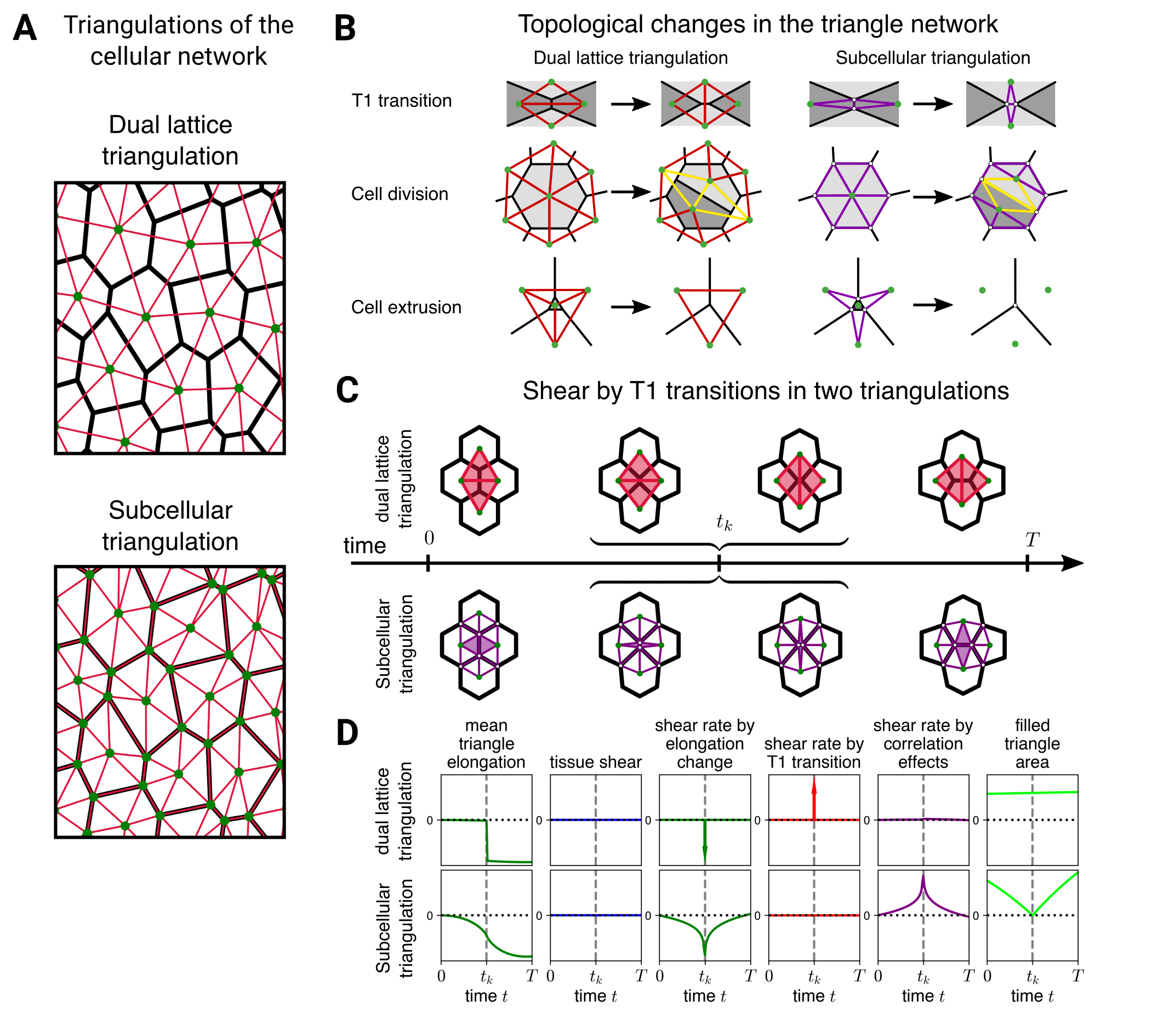}
 \caption{Quantifying cellular contributions to tissue shear using two different triangulations.
 \textbf{(A)} Definitions of the two triangulations (red) of the cellular network (black).
  Top: Dual lattice triangulation, where the cell centers (green points) of neighboring cells are connected. 
  Bottom: Subcellular triangulation, connecting each cell center with two consecutive vertices on the cell boundary.
 \textbf{(B)} Topological rearrangements in the dual lattice triangulation (left column, red triangles) and subcellular triangulation (right column, purple triangles)
 due to a T1 transition (top), cell division (middle) and cell extrusions (bottom).
 \textbf{(C)} Illustration of the shear rate contributions during a T1 transition using a dual lattice triangulation (top) or a subcellular triangulation (bottom). Between times $0$ and $t_k$, the length of the central vertical bond is imposed to shrink to zero, at which point the topology of the cell network is changed. From $t_k$ till $T$, the length is imposed to grow in the horizontal direction.
 \textbf{(D)} For the process shown in (C), from left to right, the mean cell elongation, the total shear, shear due to cell shape change, shear due to T1 transitions, shear due to correlation effects and the mean area of the colored triangles in (C). See App.~\ref{sec_subcellular_triangulation} for details.}
 \label{fig_network_dynamics}
\end{figure}

\section{Dynamics of a polygonal cell network under pure shear}

\label{sec_dynamics}

The deformation of a cellular network is quantified by the shear rate $V_0$. The overall shear of the network can be decomposed into cellular contributions. For flat triangulated networks, such a decomposition can be done exactly~\cite{etournay2015interplay, merkel2017triangles}. 
Following Refs.~\cite{etournay2015interplay, merkel2017triangles}, starting from a polygonal cellular network we can use a \textit{dual lattice} triangulation to define the shear decomposition.
We assign to each three-fold vertex of a polygon a triangle. The triangle corners are the area-weighted geometric centers of three polygons that meet at the vertex, see Fig.~\ref{fig_network_dynamics}.
This choice of triangulation of the polygonal network is however not unique. We also introduce in this paper an alternative, \textit{subcellular} triangulation. The geometry of these triangles is given by connecting the geometric center of each cell polygon to two consecutive vertices on the polygon. See Fig.~\ref{fig_network_dynamics}A for illustration.
Although the total tissue shear is the same for any triangulation, the specific choice of triangulation has an impact on the decomposition of shear.
In the following, we compare the results obtained by these two triangulations.

    \subsection{Quantifying cellular contributions to tissue deformation}

To study the dynamical response of an isotropic cellular network under shear flow, we use the stochastic vertex model introduced in Sec.~\ref{sec_vertex_model} with pure shear boundary conditions and analyze its dynamics using the two different triangulations introduced above. Following Ref.~\cite{merkel2017triangles}, the large-scale shear-rate tensor $\widetilde V_{ij}$ of the cellular network can be decomposed as: 
\begin{align}
    \widetilde V_{ij} = \frac{{\rm D} Q_{ij}}{{\rm D}t} + R_{ij} \, .
    \label{eq_shear_decomp}
\end{align}
Here and in the following, $i$ and $j$ corresponds to 2d Cartesian indices, $Q_{ij}$ is the mean cell elongation tensor and ${\rm D}/{\rm D}t$ is the corotational time derivative of a tensor (defined in Eq.~\eqref{eq_SI_hydr_model_corotational_derivative_simple_shear} of App.~\ref{sec_definitions}). The tensor $R_{ij}$ accounts for shear rate due to topological rearrangements and is a sum of four contributions:
\begin{align}
    R_{ij} = T_{ij} +  C_{ij} +  E_{ij} +  D_{ij} \, ,
    \label{eq_shear_decomp_2}
\end{align}
where the tensors $T_{ij}$, $C_{ij}$ and $E_{ij}$ account for shear rate due to T1 transitions, cell divisions and cell extrusions, respectively.
Note that we focus on cell networks where cell number is constant and such that, by good approximation, $C_{ij} = E_{ij}=0$. 
The tensor $D_{ij}$ is a shear rate due to correlations which arises when coarse-graining the single triangle shear rate over patches of triangles.
This term includes shear caused by correlations between cell rotations and cell elongation as well as correlations between cell area changes and cell elongation~\cite{etournay2015interplay,merkel2017triangles}. 

The choice of the triangulation affects how the total tissue shear rate $\widetilde V_{ij}$ is decomposed into the different cellular contributions. As we show below in a concrete example, the cell elongation contribution ${\rm D} Q_{ij}/{\rm D}t$ is only weakly affected by the choice of triangulation. However, the decomposition of $R_{ij}$ into cellular contributions (Eq.~\eqref{eq_shear_decomp_2}) depends crucially on the triangulation chosen (see Fig.~\ref{fig_network_dynamics}). In particular, the \textit{subcellular} triangulation yields by definition a vanishing T1 transition contribution:  $T_{ij}^{\rm sub}=0$, as illustrated in Panel D of Fig.~\ref{fig_network_dynamics} and demonstrated in App.~\ref{sec_subcellular_triangulation}. Compared to the dual lattice triangulation, the contribution of T1 events to the shear is transferred to the correlation term $D_{ij}$ such that the sum $T_{ij}+D_{ij}$ contributing to $R_{ij}$ remains essentially the same for both triangulations. 

Note that the trace of the velocity gradient tensor $\widetilde V_{kk}$ (summation over repeated indices is implied), which corresponds to isotropic tissue growth, can also be decomposed into cellular contributions~\cite{merkel2017triangles,popovic2017active}. Here, we only focus on the anisotropic contributions.
Finally, the tissue stress tensor $\Sigma_{ij}$ in the simulations is symmetric and can be decomposed into an isotropic pressure and a symmetric traceless part, the shear stress $\widetilde \Sigma_{ij}$.

\subsection{Shear decomposition of a polygonal network under pure shear}

\begin{figure}[t]
 \centering
 \includegraphics[width=0.95\textwidth]{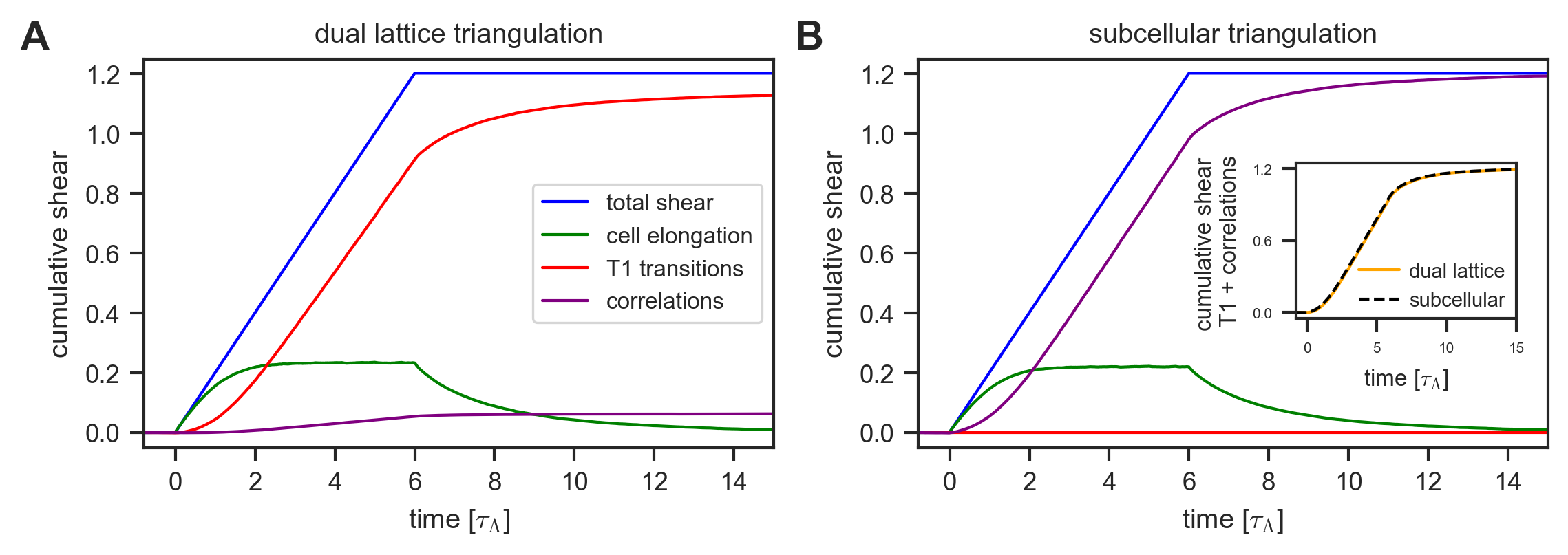}
 \caption{Dynamics of an isotropic cellular network under an imposed pure shear deformation. 
 From $t=0$ to $t=6$, we apply pure shear to the network with a rate $\widetilde{V}_{xx} = 0.2$,  after which we let the system relax while keeping the box fixed. 
 Panels show cumulatives of the total tissue shear (blue), decomposed in contributions to the shear due to changes in cell shape (green), due to T1 transitions (red) and due to correlation effects (purple). 
 \textbf{(A)} Shear decomposition using a dual lattice triangulation or \textbf{(B)} subcellular triangulation. 
 The inset in (B) shows the sum of the shear by T1 transitions and correlation effects, which gives a similar result for both triangulations.
 Vertex model parameter values used given in Table~\ref{tab_vertex_model_parameters}.
 The data is averaged over 100 realizations of the vertex model. 
 The standard error of the mean is smaller than the thickness of the curves.}
 \label{fig_isotropic_pure_shear_triangulation_comparison}
\end{figure}

We now illustrate the shear decomposition procedure in the case of a cellular network under pure shear.
Starting from an isotropic steady state of the cellular network at $t=0$, a pure shear boundary condition with constant shear rate $\widetilde V_{xx}=V_0$ along the $x$-axis is imposed on the network during a finite period, after which the box dimensions are fixed, and the system relaxes towards an isotropic steady state. In Fig.~\ref{fig_isotropic_pure_shear_triangulation_comparison}, we decompose the pure shear deformation into cellular contributions, using a dual lattice triangulation (panel A) and a subcellular triangulation (panel B).

Under pure shear, the tissue is characterized at short time by an elastic behavior as cells respond to the deformation by elongating (green curves in panels A and B). With a delay, cell elongation is then relaxed through T1 transitions (red curve in panel A), showing a viscous behavior of the cellular network at longer times. After a time $\tau_{\rm r}\simeq 2$, we observe that the shear due to cellular rearrangements (red curve in panel A, purple curve in panel B) accounts for most of the pure shear and the mean cell elongation levels off. When shearing stops at $t=6$, topological rearrangements relax the cell elongation and the tissue reaches an isotropic steady state.

Note that the viscous relaxation due to topological rearrangements is captured by different terms of the shear decomposition~\eqref{eq_shear_decomp} depending on the triangulation used. Indeed, in the subcellular triangulation, this viscous behavior is not captured by the T1 transition contribution -- which always vanishes in such triangulations -- but by the correlation contribution (purple curve in panel B). Importantly, adding the shear due to T1 transitions and the shear due to correlation effects in either of the two triangulations yields essentially the same contribution due to rearrangements, see the inset of panel B.

    \subsection{Comparison to a linear continuum model}

The viscoelastic behavior of stochastic cellular networks can be captured by a continuum model of tissues. Such a coarse-grained description does not hold at a single cell level but requires an averaging over many cells, 
as provided by the shear decomposition of a triangulated network discussed above.   
We describe in this section how  the decomposition of tissue shear is captured in a continuum model~\cite{marmottant2009role, etournay2015interplay, popovic2017active}.
In all cases below, we discuss two-dimensional tissues, but the equations can be generalized to other dimensions.

        \subsubsection{Cellular contributions to tissue deformation and tissue stress}

Similar to the shear decomposition performed for cell networks in Eq.~\eqref{eq_shear_decomp}, the anisotropic part of the deformation rate tensor $\tilde v_{ij}$ can be decomposed into cellular contributions due to changes in the mean cell elongation tensor $q_{ij}$ and shear $r_{ij}$ caused by topological rearrangements. Note that we use lower-case letters for the continuum model description. We therefore have:
\begin{equation}
 \tilde v_{ij} = \DDt{q_{ij}} + r_{ij} \, ,
 \label{eq_cont_shear_decomp}
\end{equation}
where $\DDtinline{}$ denotes the corotational derivative defined in Eq.~\eqref{eq_SI_hydr_model_corotational_derivative_simple_shear}. 
We also introduce the tissue stress $\sigma_{ij}$, which we decompose into an isotropic part and an anistropic symmetric traceless part, the tissue shear stress $\tilde \sigma_{ij}$.

        \subsubsection{Linear continuum model of an isotropic network}

\label{sec_linear_model}

We consider that the cellular network is an elastic material. To linear order, the shear stress is thus proportional to cell elongation and reads:
\begin{align}
    \tilde \sigma_{ij} \simeq \shearmod \, q_{ij} \, ,
    \label{eq_shearModulus}
\end{align}
where \shearmod{} is the shear modulus of the tissue. Moreover, 
we first consider the linear response where the rate of topological rearrangements is proportional to cell elongation and is given by the constitutive equation:
\begin{align}
 r_{ij} \simeq k_1 \, q_{ij} \, ,
 \label{eq_linear_isotropic_rij}
\end{align}
where $k_1$ is the characteristic rate of topological rearrangements, which corresponds to the inverse stress relaxation time in the linear response. As we will discuss in the following section, this linear description of the topological rearrangements given in Eq.~\eqref{eq_linear_isotropic_rij} does not fully capture the rheological properties of cellular networks when tissue shear is high and the bond tension fluctuations are low and will need to be extended.

\begin{figure}[t]
\centering
    {\includegraphics[width=0.49\textwidth]{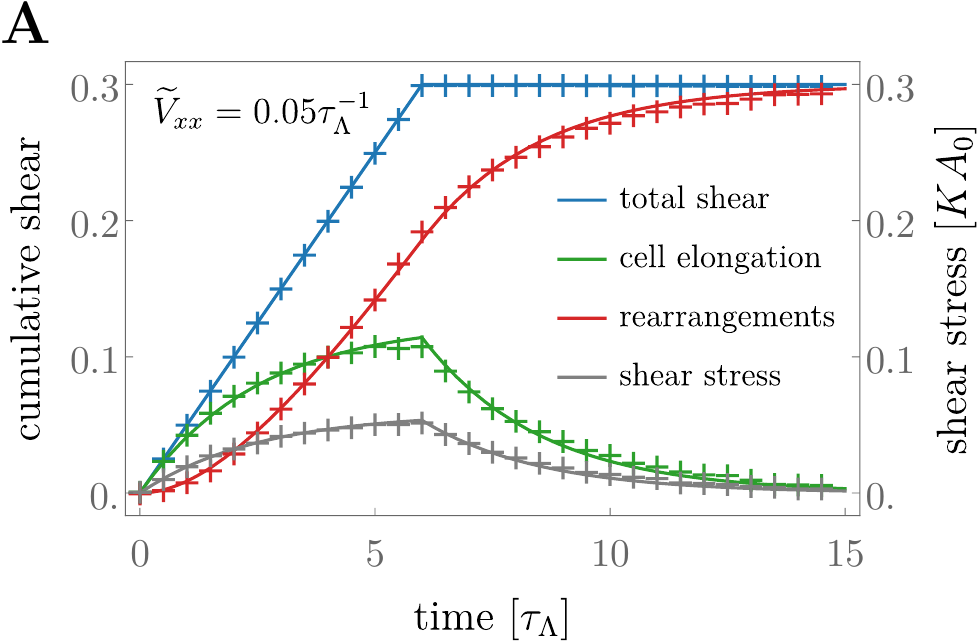}}
    \hfill
    {\includegraphics[width=0.49\textwidth]{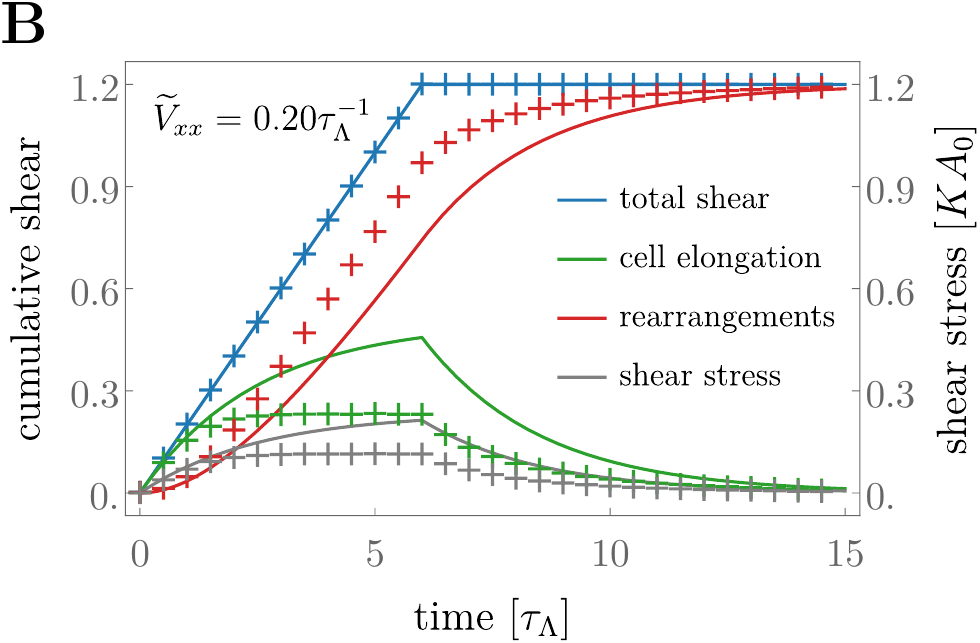}}
 \caption{%
 \textbf{(A)} Fit of the isotropic continuum model (solid lines) to the shear decomposition of the cellular network dynamics of the vertex model (crosses, data from Fig.~\ref{fig_isotropic_pure_shear_triangulation_comparison}).
 A pure shear deformation is imposed along the $x$-axis with a rate  $\widetilde V_{xx}=0.05\,\tau_\Lambda^{-1}$.
 The shear due to rearrangements in the cellular network (dark red crosses)
 is obtained by adding the shear due to T1 transitions and correlation effects.
 Fit consists of two parts: (i) deformation under imposed pure shear and (ii) relaxation back to the isotropic state,
 where we use the parameter values, $\shearmod{} = 0.47 K A_0$, $k_1 = 0.40 \,\tau_\Lambda^{-1}$.
 \textbf{(B)} Same setup as in panel A, but with a larger imposed shear rate $\widetilde V_{xx}=0.20\,\tau_\Lambda^{-1}$. The values of $\shearmod{}$ and $k_1$ of the continuum model are the same as used in panel A.
 Note that we could have obtained a good fit if we did change the parameter values.%
 }
 \label{fig_isotropic_pure_shear_hydr_model_fit}
\end{figure}

In Fig.~\ref{fig_isotropic_pure_shear_hydr_model_fit}, panel A, we compare the solution of the linear continuum model (solid lines) with the shear decomposition of the vertex model simulations (crosses).
The tissue cell elongation $q_{ij}(t)$ is obtained by solving Eqs.~\eqref{eq_cont_shear_decomp} and~\eqref{eq_linear_isotropic_rij}, together with a pure shear boundary condition setting $\tilde v_{ij}(t)$. The stress is then obtained using Eq.~\eqref{eq_shearModulus}.
The continuum model solution $q_{ij}(t)$ is given in Eq.~\eqref{eq_SI_hydr_model_isotropic_linear_pure_shear_dyn} in the Appendix and is used to fit the average cell elongation $Q_{ij}$ quantified in the vertex model simulations.
We have added the shear contributions due to T1 transitions and correlation effects ($T_{ij}+D_{ij}$) obtained from the vertex model simulations, and we compare their combined contribution to the rate of topological rearrangements  $r_{ij}$ in the continuum model.
We find that the linear continuum model with two fitting parameters $k_1$ and \shearmod{} gives an accurate description of the observed shear flows in the vertex model for this choice of parameters and shear rate.

To test the limitations of the linear model, we now compare the continuum model results to the vertex model simulations for a higher shear rate.
In the panel B of Fig.~\ref{fig_isotropic_pure_shear_hydr_model_fit}, we give the shear decomposition of the vertex model simulations  (crosses) for an applied shear rate that is four times larger compared to panel A, and we compare it with the linear continuum model results (solid lines) using the values for \shearmod{} and $k_1$ that were fit for in panel A.
The linear model now overestimates the observed mean cell elongation in the vertex model, suggesting that the values of \shearmod{} and $k_1$ would have to be changed for a better correspondence between both descriptions of the cellular network. 

In the following, we will show that this limitation of the linear model can be addressed by the description of the tissue as a \textit{nonlinear} active material. 
These nonlinearities in the vertex model are all the more pronounced when the bond tension fluctuations are low, as the system will be shown to behave more like a yield-stress material than a regular fluid~\cite{bonn2017yield,popovic2020inferring}. We therefore focus in the next section on the influence of these fluctuations on the steady-state behavior of the vertex model.

\section{Nonlinear rheology of cellular networks}

    \label{sec_non_linear_response}

In this section we continue on to explore the nonlinear properties of the vertex model. For this purpose, we investigate the  effect of line tension fluctuations $\Delta \Lambda$ (see Eq.~\eqref{eq_line_tension_dyn_isotropic}) on the rheology of cellular networks. 
At vanishing fluctuation amplitude, the vertex model has been shown to behave as a yield-stress material~\cite{popovic2020inferring}, with a viscosity that diverges as the shear rate approaches zero.
We observe that signatures of this zero-noise yielding transition can also be seen at finite values of the fluctuations and appear in terms of nonlinear rheology of the model tissue. Even at large noise strength, where the vertex model could be approximately described as a Newtonian fluid (see previous section), we observe a shear-thinning behavior at larger shear rates.
As we show below, these rheological properties of the vertex model can be captured by a continuum model with nonlinear constitutive equations.

\begin{figure}[t]
\hfill
{\includegraphics[width=0.48\textwidth]{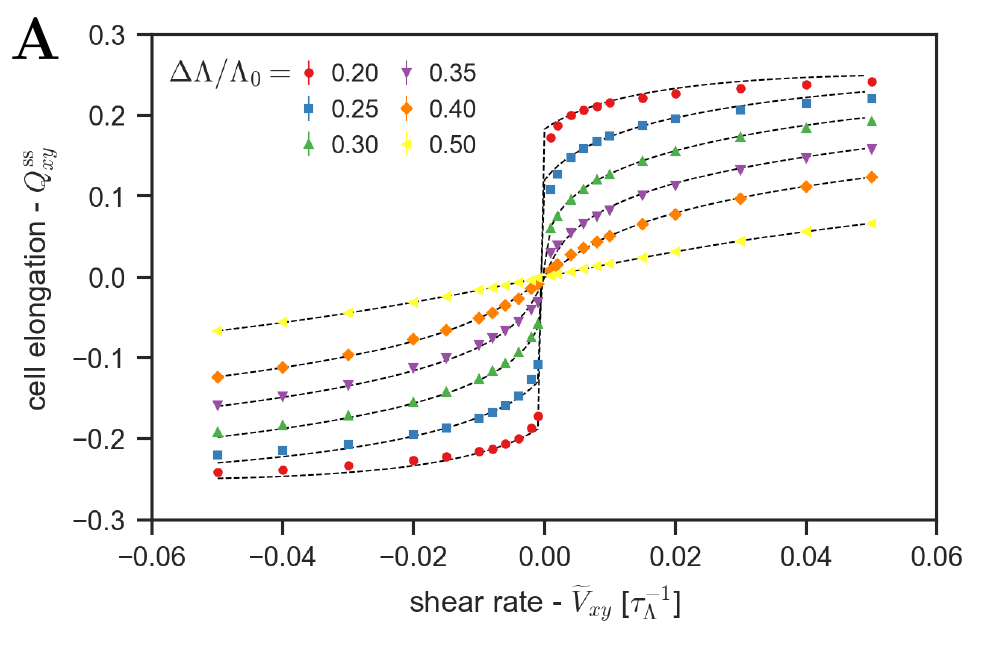}}
{\includegraphics[width=0.48\textwidth]{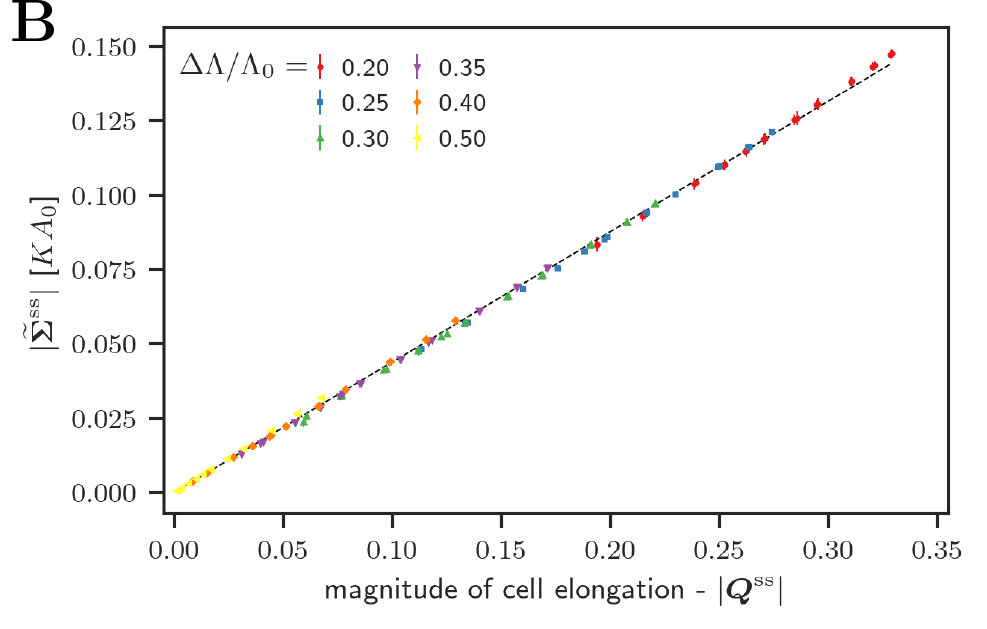}}
\hfill
 \caption{
 Nonlinear response of the isotropic cellular network to different rates of imposed simple shear deformation.
 The color of the symbols indicate different values used for the magnitude of the line tension fluctuations in the simulations, $\Delta\Lambda$, which are given in the legend.
 \textbf{(A)} Steady state value of the $xy$-component of the mean cell elongation tensor $Q^{\rm ss}_{xy}$ against the imposed simple shear rate, $\widetilde{V}_{xy}$.
 Dashed black lines show separate fits of the nonlinear continuum model in Eq.~\eqref{eq_hydr_model_non_linear_isotropic} for each value of $\Delta\Lambda$.
 Fitted values for $k_1$ and $k_3$ are given in Table~\ref{tab_hydr_model_isotropic_shear_mod}. 
 The $xx$-component of the mean elongation tensor are given in Fig.~\ref{fig_SI_isotropic_simple_shear_vary_bond_tensions}.
 \textbf{(B)} Magnitude of the steady-state shear-stress tensor against the magnitude of the mean cell elongation. The color of the symbols correspond to the values of $\Delta\Lambda$. Different symbols of the same color correspond to different applied shear rates. 
 The dashed line corresponds to a linear fit (including the origin) through all the data points. The tissue elastic shear modulus $\mu_\mathrm{s} = 0.44$ is given by the slope of this linear fit. 
}
 \label{fig_isotropic_simple_shear_vary_bond_tensions}
\end{figure}


    \subsection{Shear thinning and nonlinear rheology in an isotropic polygonal cell network}

Here we characterize the effect of line tension fluctuations on the shear flow in an isotropic cellular network. To this end, we study for different magnitudes of the line tension fluctuations~$\Delta \Lambda$ the dependence of the steady-state cell elongation on simple shear.

In Fig.~\ref{fig_isotropic_simple_shear_vary_bond_tensions}, panel A, we display the response of the steady-state mean cell elongation $Q^{\rm ss}_{xy}$ to the imposed rate of a simple shear deformation $\widetilde V_{xy}$, for different values of the magnitude of line tension fluctuations $\Delta\Lambda$. Filled symbols correspond to vertex model simulations and dotted lines correspond to a fit of a nonlinear model discussed below. We emphasize that the magnitude of the steady-state tissue shear stress, $ |\bm{\widetilde\Sigma}^{\rm ss}| = \sqrt{(\widetilde \Sigma_{xx}^{\rm ss})^2 + (\widetilde \Sigma_{xy}^{\rm ss})^2}$, is directly proportional to the magnitude of the steady-state mean cell elongation $|\bm{Q}^{\rm ss}| = \sqrt{(Q_{xx}^{\rm ss})^2 + (Q_{xy}^{\rm ss})^2}$, see panel B. As a consequence, panel A can be read directly as a shear stress versus shear rate plot, from which the rheological properties of cell networks can be discussed. 

A first observation is shear thinning: the viscosity of the cellular network is lower at higher shear rates, since the slope of the response of tissue mean cell elongation to the applied shear rate decreases at higher shear rates (see Fig.~\ref{fig_isotropic_simple_shear_vary_bond_tensions}A). 
This can be understood qualitatively by observing that to accommodate an imposed shear rate, cells in the network elongate but more importantly must undergo T1 transitions. 

As the shear rate increases, the rate of T1 transitions increases because additional transitions are driven by the extra shear. These T1 transitions keep the cell elongation at a smaller level. Since the tissue stress is proportional to cell elongation, a shear-thinning behavior follows.

Second, our results suggest a stronger nonlinear behavior at low noise magnitude, where the signature of the yielding transition that exists in the vertex model at vanishing noise amplitude~\cite{popovic2020inferring} is more apparent. We observe a glassy behavior in this regime, indicated by a large viscosity at small values of the imposed shear rate, which is then followed by a shear-thinning a larger shear rates (see Fig.~\ref{fig_isotropic_simple_shear_vary_bond_tensions}).

\subsection{Nonlinear continuum model of an isotropic network}

\begin{figure}[t]
\centering
    {\includegraphics[width=0.49\textwidth]{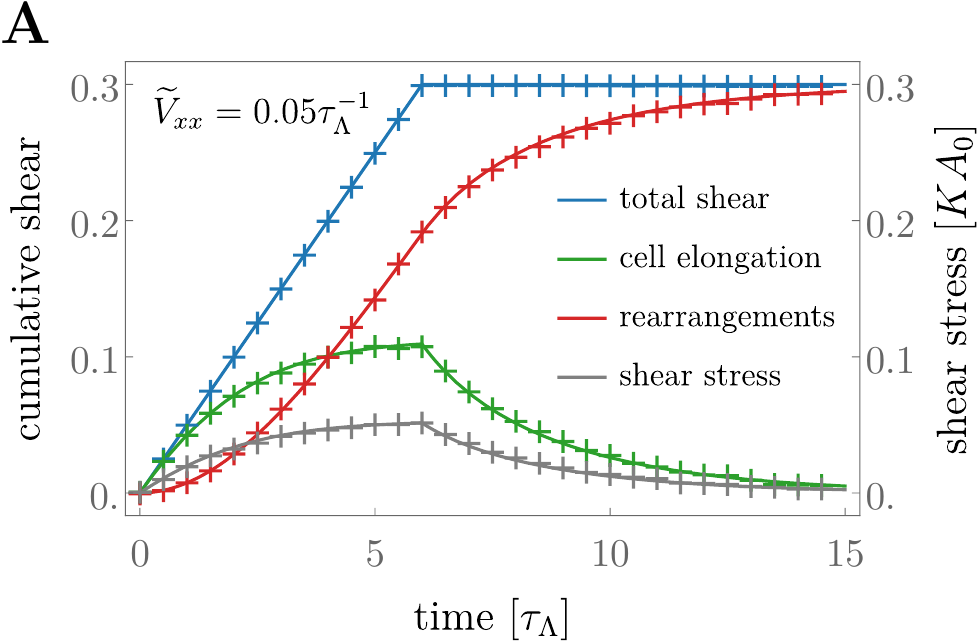}}
    \hfill
    {\includegraphics[width=0.49\textwidth]{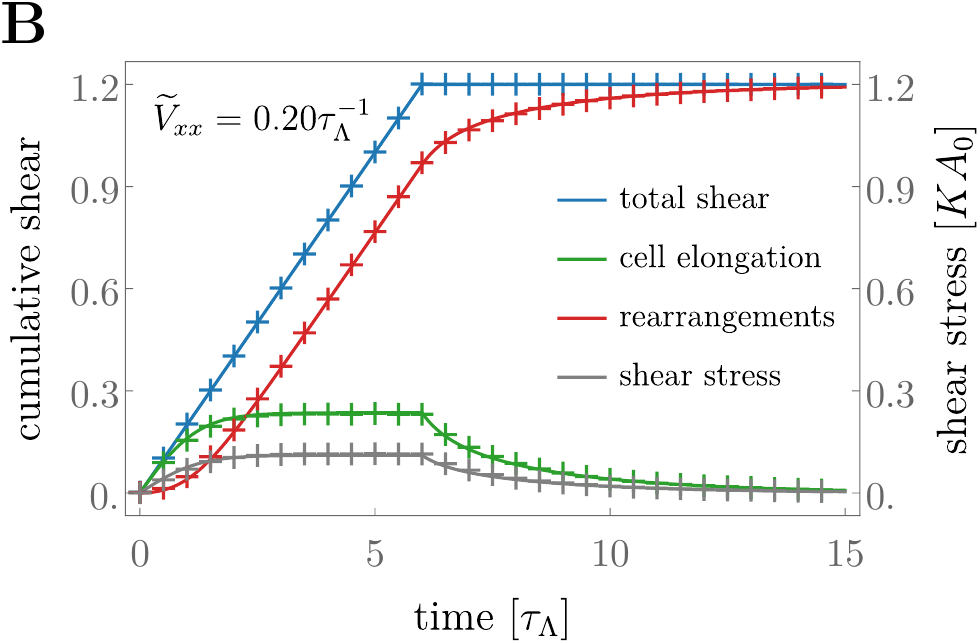}}
 \caption{
     Nonlinear fit (solid lines) to the shear decomposition of the cellular network dynamics of the vertex model (crosses, data from Fig.~\ref{fig_isotropic_pure_shear_triangulation_comparison}).
     \textbf{(A)}  A pure shear deformation is imposed along the $x$-axis with a rate $\widetilde V_{xx}=0.05\,\tau_\Lambda^{-1}$.
     \textbf{(B)} Same setup as in panel A, but with a larger imposed shear rate $\widetilde V_{xx}=0.2\,\tau_\Lambda^{-1}$.
     The same fit parameters $k_1=0.32$ and $k_3=4.72$ have been used for both panels, hence solving the limitation of the linear model displayed in Fig.~\ref{fig_isotropic_pure_shear_hydr_model_fit}. Note that the best fit for the shear stress (gray curves) is obtained with $\shearmod{} = 0.47 K A_0$, which is the same value as in Fig.~\ref{fig_isotropic_pure_shear_hydr_model_fit}.
}
 \label{fig_nonlinear_fit}
\end{figure}

We have discussed in Sec.~\ref{sec_linear_model} a description of tissues with material properties that depend linearly on the tissue state (see Eqs.~\eqref{eq_shearModulus} and~\eqref{eq_linear_isotropic_rij}). This linear model predicts a linear response of the mean cell elongation to small values of the applied shear rate and a vanishing cell elongation as the shear rate approaches zero, see App.~\ref{sec_hydr_model_linear_isotropic_shear}. However, such a linear model does not agree with what we observe in the vertex model simulation of the cellular network, as discussed above and shown in Fig.~\ref{fig_isotropic_simple_shear_vary_bond_tensions}, especially for lower values of the magnitude of the line tension fluctuations. In the low line tension fluctuations regime, the linear elastic behavior of the tissue stress as a function of cell elongation, $\tilde\sigma_{ij}=\mu_{\rm s} \, q_{ij}$, is expected to remain valid, as suggested by panel B of Fig.~\ref{fig_isotropic_simple_shear_vary_bond_tensions}. However, the fluctuation magnitude is expected to play a crucial role in the rate of topological rearrangements: at low magnitude, rearrangements triggered by bond tension fluctuations occur extremely rarely, inducing the appearance of a glassy regime at vanishing shear rate.

Therefore, we expand the linear constitutive equation for the rate of rearrangements tensor $r_{ij}$, with the lowest-order nonlinear dependence on the cell elongation tensor allowed by the symmetry of the system. Using the fact that $r_{ij}$ is a symmetric traceless tensor, we thus write the nonlinear constitutive equation:
\begin{align}
    r_{ij} = k_1 \, q_{ij} + \frac{k_3}{2}\mathrm{Tr}\left[\bm{q}\cdot\bm{q}\right]q_{ij} \, ,
 \label{eq_hydr_model_non_linear_isotropic}
\end{align}
where $k_3$ is a rate for the third-order dependence of $r_{ij}$ on cell elongation. Note that this simple form is obtained in two dimensions, where the square of a symmetric traceless tensor is always proportional to the identity: ${\bm q}^2 = q^2 \mathbb{1}$, where $q=\sqrt{q_{xx}^2+q_{yy}^2}$ is the norm of the nematic tensor. In three dimensions, an additional term would need to be considered. The results shown in Fig.~\ref{fig_isotropic_simple_shear_vary_bond_tensions}B reveal that Eq.~\eqref{eq_shearModulus} does not need to be extended to nonlinear order.

The nonlinear constitutive equation~\eqref{eq_hydr_model_non_linear_isotropic} can first be used to obtain the dynamics of the cell elongation and cell rearrangements of a cell network under pure shear. The nonlinear fit to the same data as in Fig.~\ref{fig_isotropic_pure_shear_hydr_model_fit} leads to an excellent agreement with the simulation results using only two fit parameters that are the same for both shear rates, see Fig.~\ref{fig_nonlinear_fit}. In addition to providing a better agreement with the data, the nonlinear model solves the limitation of the linear model, which did not provide a satisfying fit for two different shear rates if the same parameters where used.

We then use the nonlinear constitutive equation~\eqref{eq_hydr_model_non_linear_isotropic} to obtain the behavior of a cell network under simple shear. We can in particular determine the steady-state cell elongation $q_{ij}^{\rm ss}$ in the network as a function of the imposed shear rate $v_0$, as presented in App.~\ref{sec_hydr_model_non_linear_isotropic_shear}. We fit the resulting expression for $q_{ij}^{\rm ss}$ to the simulation data given in Fig.~\ref{fig_isotropic_simple_shear_vary_bond_tensions}.
We find an excellent agreement between the nonlinear continuum model and the simulation data, both at high and low levels of line tension fluctuations. 

At low level of line tension fluctuations and for small shear rates, the network exhibits a glassy behavior which arises when the inverse shear rate becomes slower or comparable to the stress relaxation time in the linear response.
In our continuum model, this glassy regime is only partially captured and is reflected by a negative value of the coefficient $k_1$, in which case an \textit{apparent} yield stress arises (see Fig.~\ref{fig_isotropic_simple_shear_vary_bond_tensions}). 
Indeed, as discussed in App.~\ref{sec_hydr_model_non_linear_isotropic_shear}, the steady-state elongation has the following behavior at low shear rate:
\begin{align}
    q_{xy}^{\rm ss} =
    \begin{cases} \frac{v_0}{k_1} + \mathcal{O}(v_0^2) &\mbox{if } k_1\geq0 \, , \\
    \pm \frac{\sqrt{-k_1(2 k_1+k_3)}}{\sqrt{2}\, k_3}-\left(\frac{2}{k_3}+\frac{1}{2 k_1}\right) v_0+ \mathcal{O}(v_0^2) &\mbox{if } k_1<0 \, ,
    \end{cases} 
\end{align}%
where we can identify the yield stress $\sigma^{\rm y}_{xy} = \pm \mu_{\rm s} \sqrt{-k_1(2 k_1+k_3)}/\sqrt{2}\, k_3$ (the minus sign corresponds to the negative shear-rate branch) whenever $k_1$ is negative.

Negative $k_1$ are found by our fits to the simulation data in situations where the rheology is qualitatively captured by a yield stress.
Such an apparent yielding behavior is observed below $\Delta\Lambda_{\rm c} \simeq 0.3$.
However, in the presence of a finite amount of fluctuations, a very long simulation would reveal that the true yield stress is actually zero. 
To fully capture the glassy regime at low line tension fluctuations requires a more detailed analysis that is beyond the scope of the present paper~\cite{popovic2020thermally,matoz-fernandez2017nonlinear}.

\subsection{Rheology of a network with anisotropic active cell stress}

So far we have discussed the nonlinear properties of an isotropic cellular network, in which the ground state is a regular hexagonal network, without a preferred axis. We now discuss the rheology of an anisotropic tissue. The anisotropy of the network is associated with a nematic
\begin{align}
  \mathcal{P}_{ij} = 2 P_i P_j - \delta_{ij} \, ,
\end{align}
assigned to each polygon,
which is constructed from a unit polarity vector $P_i$, motivated by planar cell polarity~\cite{wang2007tissue}. In two dimensions, this nematic can be parameterized by a single angle~$\Psi$ which defines the direction of the anisotropy axis (see App.~\ref{sec_nematic_tensor_decomposition} for details). 
Although the polarity field may change with time and may depend on cell stress, we consider for simplicity that $\Psi$ is constant in the following.
Moreover, this nematic field gives rise to anisotropic active stresses in the tissue. In the vertex model, this is implemented by adding the work performed by this active stress $\Sigma^{\mathrm{a}}_{ij}$ to the vertex model work function, which reads:
\begin{align}
  W = W_0 - \sum_\alpha \frac{1}{2}A^\alpha \Sigma^{\mathrm{a}}_{ij} \shearQVM{}^\alpha \, ,
 \label{eq_tissue_work_anisotropic}
\end{align}
where $W_0$ is defined in Eq.~\eqref{eq_tissue_work} and $\shearQVM{}^\alpha$ is the cell shape tensor of each cell $\alpha$
\begin{align}
\bm G^\alpha = \frac{1}{A^\alpha}\sum_{\langle m,n \rangle} \bm{\widetilde{\mathcal{L}}}_{mn} \, .
 \label{cell_shape_tensor}
\end{align}
Here, $\bm{\widetilde{\mathcal{L}}}_{mn}$ is a symmetric traceless tensor defined by the orientation of the bond connecting vertices $m$ and $n$. From the vector $\bm{\mathcal{L}}_{mn}$ pointing from vertex $m$ to vertex $n$, we define \mbox{$\bm{\widetilde{\mathcal{L}}}_{mn}=\bm{\mathcal{L}}_{mn} \otimes \bm{\mathcal{L}}_{mn} - \tfrac{1}{2}\mathbb{1}$}
where $\otimes$ is the tensor product. 
The cell shape tensor $\shearQVM{}^\alpha$ quantifies the deviation of the cell shape from isotropic shapes, for which $\shearQVM{}^\alpha=0$.
See Fig.~\ref{fig_polygonal_cell_network} for an illustration.
The direction of the anisotropic active stress is set by the cell nematic as $\Sigma^{\mathrm{a}}_{ij} = \Sigma^{\mathrm{a}} \mathcal{P}_{ij}$, where $\Sigma^{\mathrm{a}}$ is the magnitude of the active stress. 
Note that anisotropy in the system in general could also be caused by anisotropic bond tensions and bond tension fluctuations. This will be the subject of a separate paper~\cite{inpreparation}.

To characterize the effects of active stress on the shear flow in cellular networks, we study the response of the mean cell elongation in the network to the rate of imposed pure shear. We vary the angle~$\Psi$ of the polarity axis $\mathcal{P}_{ij}$, while the direction of the imposed pure shear is kept along the $x$-axis. 
Then, starting from a steady-state network configuration under a fixed boundary condition, we apply a pure shear deformation along the $x$-axis, until the magnitude of the cell elongation tensor has plateaued, and then quantify the mean cell elongation tensor $Q^{\rm ss}_{ij}$. 
We characterize this mean cell elongation tensor by its magnitude $|\bm Q^\mathrm{ss}|$ and the angle of orientation of its axis $\Phi^\mathrm{ss}$ (see App.~\ref{sec_nematic_tensor_decomposition}). As a consequence of the active anisotropic cell stress, the steady-state mean cell elongation tensor $Q^{\rm ss}_{ij}$ does not vanish even when the external shear is zero. To discriminate between the effect of the applied pure shear deformation and that of the active stress field, we define the norm $|\bm{Q}^\mathrm{ss}_0|$ and orientation angle $\Phi^\mathrm{ss}_0=\Psi$ of the elongation tensor for a vanishing external shear.

In Fig.~\ref{fig_active_stress_polarity_dependence_ss_fit}, panels A and B, we show the perturbation of the magnitude $|\bm Q^\mathrm{ss}|-|\bm Q^\mathrm{ss}_0|$, and the perturbation of the axis direction $\Phi^\mathrm{ss} - \Psi$ of the mean elongation tensor due to the applied shear.
The results clearly show that the response of a cellular network to shear strongly depends on the angle between the applied deformation and the active stress.
Under a nonzero pure shear rate the magnitude of cell elongation (panel A) has approximately a cosine dependence on the polarity angle, with strongest changes in the magnitude of elongation at the angles $\Psi=0$ and $\Psi=\pi/2$.
The change in the angle of the elongation tensor due to the pure shear has approximately a sinusoidal dependence on $\Psi$, with maximal perturbations at $\Psi=\pi/4$ and $\Psi=3\pi/4$.
Note that for a shear rate $\widetilde{V}_{xx}=0.05$, the magnitude of elongation at $\Psi=0$ increases by about $0.06$, while the magnitude at $\Psi=\pi/2$ decreases by about $-0.08$, revealing that the response is not simply a cosine.
This is a signature of a nonlinear response of the cellular network to the applied shear.
To highlight this nonlinear response, we show in panel C of Fig.~\ref{fig_active_stress_polarity_dependence_ss_fit} the change in the magnitude of the mean cell elongation due to a small change in the imposed shear rate, evaluated at $\widetilde{V}_{xx}=0.025 \tau^{-1}_\Lambda$.
We find that the mean cell elongation in the tissue is more sensitive to changes in the shear rate when cells are elongated in a direction perpendicular to the axis of pure shear ($\Psi=\pi/2$) as compared to the case when cells are elongated parallel to the axis of pure shear ($\Psi=0$).

\subsection{Nonlinear continuum model of network with active stress}

\label{sec_nonlinear_model_anisotropy}

\begin{figure}[t]
    \centering
    \includegraphics[width=\textwidth]{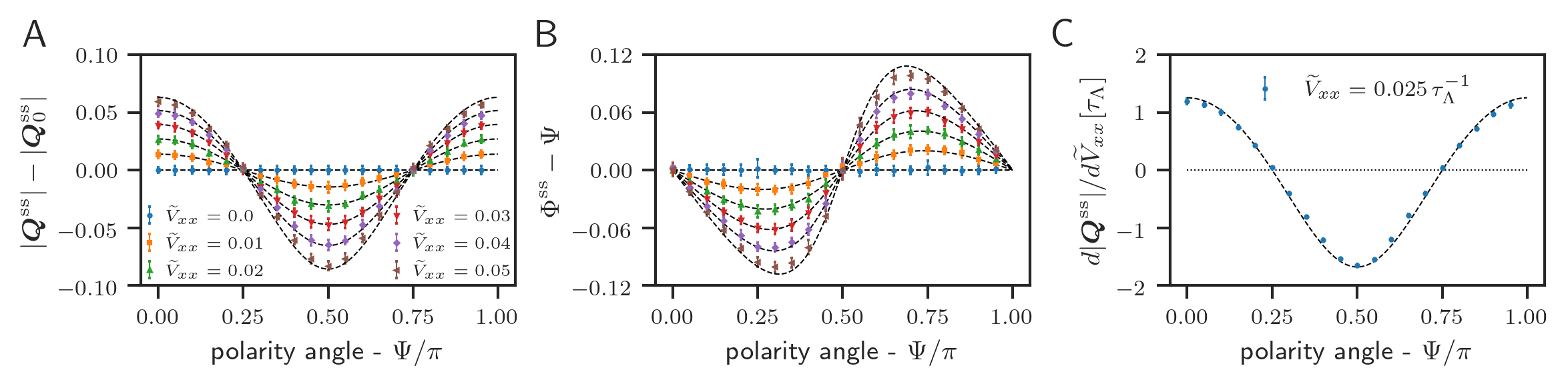}
    \caption{
    Changes in the steady state mean cell elongation due to an imposed pure shear deformation in a cellular network with anisotropic cell stress, against the angle of the polarity axis, $\Psi$.
    In panels A and B, different colors correspond to different applied pure shear rates, given in the legend of panel A. Dotted lines show the fit of the nonlinear anisotropic model with the constitutive equation~\eqref{eq_hydr_model_anisotropic_stress_non_linear}.
    \textbf{(A)} Change in the magnitude of the mean cell elongation due to pure shear $|\bm {Q}^\mathrm{ss}|$, compared to the case when no shear is imposed, $|\bm Q^\mathrm{ss}_0|$.
    \textbf{(B)} Change in the angle $\Phi^\mathrm{ss}$ of the mean elongation tensor compared to the angle~$\Psi$ when no shear is imposed.
    \textbf{(C)} Susceptibility of the magnitude of cell elongation to changes in the applied shear rate. Derivative taken at $\widetilde{V}_{xx}=0.025\tau_\Lambda^{-1}$.
    Dotted lines give the prediction obtained from the nonlinear model of Sec.~\ref{sec_nonlinear_model_anisotropy}.
    Parameter values obtained from the fit given in Table~\ref{tab_SI_hydr_model_anisotropic_stress_fitted_values}. }
    \label{fig_active_stress_polarity_dependence_ss_fit}
\end{figure}

To capture the nonlinear response of the network to a combination of external pure shear and active cell stress presented in Fig.~\ref{fig_active_stress_polarity_dependence_ss_fit}, we extend the isotropic constitutive equation for the rate of rearrangements $r_{ij}$ as given by Eq.~\eqref{eq_hydr_model_non_linear_isotropic} to the case of an anisotropic material. This anisotropy is captured by the nematic field $p_{ij}$, which is the continuum model equivalent of $\mathcal{P}_{ij}$. In the presence of this additional anisotropy, the most general constitutive equation for the traceless symmetric tensor $r_{ij}$ in two dimensions reads:
\begin{align}
\begin{split}
    r_{ij} &= 
    \left(k_1 + \frac{k_2}{2}\mathrm{Tr}[\bm p\cdot \bm q] \right) q_{ij} 
    + \left( \gamma_0 + \frac{\gamma_1}{2}\mathrm{Tr}[\bm p\cdot \bm q] \right) p_{ik} q_{kl} p_{lj}
    +  \delta_0 \, p_{ik}q_{kl}p_{lm}q_{mn}p_{nj} \\
    &+ \left( \lambda_0 + \frac{\lambda_1}{2}\mathrm{Tr}[\bm p\cdot \bm q] 
    + \frac{\lambda_2}{2} \mathrm{Tr}[\bm q\cdot \bm q] 
    + \frac{\lambda_3}{2}\mathrm{Tr}[\bm p\cdot \bm q \cdot \bm p\cdot \bm q] 
    + \frac{\lambda_5}{4}\mathrm{Tr}[\bm p\cdot \bm q]^2 \right) p_{ij}
    + \mathcal{O}(|\bm q|^3 )\, ,
\end{split}\label{eq_hydr_model_anisotropic_stress_non_linear}
\end{align}
where we have considered the lowest nonlinear terms in the cell elongation $\bm q$.
Importantly, note that these lowest-order nonlinear terms are of order two in ${\bm q}$, whereas they were of order three in the isotropic case (see Eq.~\eqref{eq_hydr_model_non_linear_isotropic}). Contributions of order two do not exist in the isotropic case because $\bm q^2 \propto \mathbb{1}$ (in spatial dimension $d=2$), while they do exist in the presence of the anisotropic order parameter $\bm p$. In addition, we also need to add linear anisotropic terms to the stress constitutive equation~\eqref{eq_shearModulus}, see App.~\ref{sec_SI_anisotropic_stress}.

In order to obtain a closed expression for $q_{ij}$ and to fit it to the mean cell elongation $Q_{ij}$ obtained from the vertex model simulations, we parameterize the elongation tensor by its norm $q=|\bm q|$ and orientation angle $\phi$ (see App.~\ref{sec_nematic_tensor_decomposition}).
Writing $\phi=\psi+\delta\phi$, we expand the elongation tensor in $\delta\phi$. To linear order in $\delta\phi$, Eq.~\eqref{eq_hydr_model_anisotropic_stress_non_linear} then reads:
\begin{equation}
    r_{ij}^{(1)} = \left[ g(q)\,\delta_{ik} + 2\delta\phi \, h(q)\, \varepsilon_{ik} \right] p_{kj} + \mathcal{O}(\delta\phi^2) \, ,
    \label{eq_hydr_model_anisotropic_stress_non_linear_lin_phi}
\end{equation}
where $\varepsilon_{ij}$ is the fully antisymmetric tensor in two dimensions with $\varepsilon_{xy} = -1$, $\varepsilon_{yx}=1$ and $\varepsilon_{xx}=\varepsilon_{yy}=0$. Here we have introduced $g(q)=(\beta_2 q^2/2 + \beta_1 q + \lambda_0)$ and $h(q)=(\beta_4 q^2+\beta_3 q +\lambda_0)$, where the effective parameters $\beta_{1,2,3,4}$ are given in Eq.~\eqref{eq_SI_hydr_non_linear_anisotropic_parameters} of App.~\ref{sec_SI_anisotropic}. We can now solve the steady-state shear decomposition with pure shear boundary conditions, $\tilde v_{ij} = r_{ij}^{(1)}$, to obtain a closed expressions for $q$ and $\delta\phi$, see Eq.~\eqref{eq_SI_hydr_model_anisotropic_stress_linearized2_pure_shear}. 

Figure~\ref{fig_active_stress_polarity_dependence_ss_fit}, panels A and B, shows the fits of this continuum model to the vertex model data using Eq.~\eqref{eq_SI_hydr_model_anisotropic_stress_linearized2_pure_shear}. Values of the fit parameters are given in  Table~\ref{tab_SI_hydr_model_anisotropic_stress_fitted_values}. We find that the continuum theory agrees well with the data.
This shows that the continuum model correctly captures the nonlinearities of the magnitude of mean cell elongation in response to changes in the shear rate, as shown in panel C.

\section{Discussion and conclusion}

We have discussed the nonlinear rheology of cellular networks described by a vertex model. We have analyzed the rheology using a shear decomposition based on a triangulation method~\cite{merkel2017triangles}, and have shown that the nonlinear properties of the cell network can be captured by a continuum model.

We have used two different triangulation schemes (dual lattice and subcellular) to show that these different choices lead to different definitions of contributions to shear by T1 transitions and by correlations. We show that the sum of both contributions, which we call the contribution from cell rearrangements, is independent of the choice of triangulation. The choice of triangulation cannot have a meaning in the continuum limit, and consistent with that fact we have shown that the cell rearrangement contribution is well-captured by the continuum theory.

The rheology of the cellular network is strongly governed by fluctuations. We have focused on a vertex model with bond tension fluctuations.
Bond tension fluctuations lead to fluctuations of bond length, which can in turn trigger stochastic topological rearrangements in the network that can relax elastic stresses.
As a result, the network can exhibit viscoelastic behavior with a characteristic relaxation time.
For increasing shear rates, we find a nonlinear response in the steady-state stress which corresponds to shear thinning, and is well-captured by a cubic term in the continuum theory.
At low fluctuation strength, nonlinear effects become strong, and the system exhibits a glassy behavior with large viscosity at small shear rate.
This nonlinear behavior of cell networks is a signature of the zero-noise yielding transition that is observed in the vertex model~\cite{popovic2020inferring} and that remains pronounced at finite noise strength. In our mean-field continuum theory, this behavior is reflected by an apparent yield stress below a characteristic noise strength.
Since the contributions from cell divisions and extrusions are small in our model, the shear-thinning and other nonlinear properties that we report rely essentially on T1 transitions. However, it has been shown that cell divisions and extrusions can also fluidify a tissue and cause shear thinning~\cite{ranft2010fluidization,matoz-fernandez2017nonlinear}.
Hence, studying the combined effect of these different sources of cellular fluctuations will be especially relevant for biological tissues, where all these cellular events contribute significantly to the macroscopic rheological properties~\cite{guirao2015unified,etournay2015interplay}.

It will be an interesting challenge to relate the coefficients in the continuum theory to the parameters of the vertex model.
One promising approach would be to derive equations for the distribution of bond lengths in the tissue in the presence of external shear and relate those to the macroscopic network behavior. Such approaches have recently been shown to be a valuable tool to investigate the rheological properties of the vertex model in the noiseless regime~\cite{popovic2020inferring}. Relating such approaches to experimentally determined cell bond length distribution in tissues could help understand the rheological properties of real tissues. 

Nonlinearities can be systematically taken into account in the continuum theory based on symmetry arguments. This allowed us to generalize our approach to discuss the rheology of anisotropic cell networks.
At the vertex model level, we have focused on an anisotropy that stems from a preferred axis, along which cells can exert active stresses.
In addition, cell bond tension can be anisotropic, which can be studied in the same framework~\cite{inpreparation}.
In the presence of anisotropy, the rheological response depends on the angle between the axis of shear and the axis of anisotropy and exhibits interesting nonlinearities. We have shown that these features are  well-captured by the nonlinear continuum theory.
The consideration of such anisotropy is a first step to build an understanding of biological tissues, in which chemical signals such as planar cell polarity pathways~\cite{wang2007tissue} can create large-scale patterns of tissue polarity, which can influence their mechanical and rheological properties, for instance by biasing the axis of cell division~\cite{gho1998frizzled}.
We have considered in our vertex model a \textit{global} nematic anisotropy. In the future, it will be important to discuss the emergence of self-organization of such polarity via \textit{local} signals~\cite{aigouy2010cell,sagner2012establishment}.
Furthermore, spatially separated patches of cells have in general different mechanical properties and may present different anisotropy axes~\cite{jain2020regionalized}. This is the case, for example, in the \textit{Drosophila} wing disc pouch, where cells along the dorso-ventral axis are on average smaller and less elongated~\cite{dye2017cell,dye2020selforganized}. Investigating the role of spatial dependence of tissue properties in developmental processes is an important challenge for future works.

\subsection*{Acknowledgements}

We thank Marko Popović for enlightening discussions and for his useful comments on the manuscript. M.M.I. acknowledges funding from Science and Engineering Research Board (MTR/2020/000605) and the hospitality at MPI-PKS, Dresden.

\newpage
\appendix

\section{Vertex model simulations}
\label{sec_vertex_model_implementation}
In this appendix we give a more detailed description of the vertex model simulations.

\begin{table}[b]
  \begin{tabular}{ |c|l|r|c| }
    \hline
    \multicolumn{4}{|c|}{vertex model parameters} \\
    \hline
    \textbf{symbol} & \textbf{explanation} & \textbf{value} & \textbf{unit} \\
    \hline
    \multicolumn{4}{|c|}{mechanics} \\
    \hline
    $\Lambda_0$         & mean line tension    & 0.12 & $K A_0^{3/2}$ \\ 
    $\Gamma$            & perimeter elasticity & 0.04 & $K A_0$ \\ 
    $\Sigma^\mathrm{a}$ & active cell stress magnitude & 0.04 & $K A_0$ \\ 
    \hline
    \multicolumn{4}{|c|}{dynamics} \\
    \hline
    $\Delta\Lambda$     & line tension fluctuations magnitude & 0.06 & $K A_0^{3/2}$ \\ 
    $\delta t$          & time step for numerical integration & 0.01 & $\tau_\Lambda$ \\ 
    \hline
  \end{tabular}
  \caption{Parameter values used in the simulations of the vertex model, expressed in dimensionless units. 
  In case different values are used for simulation results, it is stated in the caption of the figure showing the results.}
  \label{tab_vertex_model_parameters}
\end{table}

\subsection{Dimensionless parameters}
In our simulations of the vertex model we use parameters expressed in dimensionless units.
To this end, we will choose $\tau_\Lambda$ as the typical timescale, $A_0^{1/2}$ as the typical length scale and $K A_0^2$ as the typical energy scale in our model.
The effective parameters are thus reduced to \mbox{$\bar{\Lambda}_{mn}=\Lambda_0/(K A_0^{3/2})$}, \mbox{$\bar{\Gamma}=\Gamma/(K A_0)$}, \mbox{$\Delta\bar{\Lambda}=\Delta\Lambda/(K A_0^{3/2})$}.
The parameters $\tau_\Lambda$, $K$ and $A_0$ are unity in these units.
In the rest of the paper we use dimensionless units and omit the bar on the parameter symbols for simplicity.
We choose the parameter values $\Lambda_0=0.12$ and $\Gamma=0.04$ which are known to produce network configurations that agree well with those observed in the wing disk epithelium \cite{farhadifar2007influence}.
Other parameter values are listed in Table~\ref{tab_vertex_model_parameters}.

\subsection{Model initialization}

Each simulation is initialized as a network of $N_x$ by $N_y$ regular hexagonal cells, with box dimensions $L_x$ and $L_y$ set such that the network work function Eq.~\eqref{eq_tissue_work}, is in its ground state \cite{farhadifar2007influence}. 
We first propagate the system with a high magnitude of line tension fluctuations, $\Delta\Lambda/\Lambda_0=0.5$, until the system has reached a steady state configuration at $t=50\tau_\Lambda$. Once the system is prepared, we distinguish the following three cases.

(i) In case of a system under pure shear, presented in Figs.~\ref{fig_isotropic_pure_shear_triangulation_comparison} and~\ref{fig_isotropic_pure_shear_hydr_model_fit}, we initialize the system with $N_x=10$ and $N_y=40$ cells. After reaching steady state under a fixed boundary condition, we start deforming the simulation box with a pure shear with a rate $\tilde V_{xx}=0.2$ and $0.05\,\tau_\Lambda^{-1}$ for a time $6\,\tau_\Lambda$. After which we stop the pure shear and let the system relax under a fixed box boundary condition.

(ii) In case of a system under simple shear, presented in Fig.~\ref{fig_isotropic_simple_shear_vary_bond_tensions}, the cell network size is $30\times30$.
After system preparation, we activate the simple shear with the required rate and set the magnitude of the line tension fluctuations $\Delta\Lambda$ to the required value. 
We then let the system run until the mean cell elongation of the system has plateaued.
Depending on the magnitude of the line tension fluctuations, the time for the mean cell elongation to reach its plateau value ranges from 50 to  $10^4\tau_\Lambda$.
After the mean cell elongation has reached plateau, we measure its time averaged value over 100$\tau_\Lambda$ in the simulation.
This gives the reported steady-state values of the mean cell elongation under simple shear.

(iii) In case of a system with anisotropic active cell stress under pure shear, presented in Fig.~\ref{fig_active_stress_polarity_dependence_ss_fit}, we initialize the system with $N_x=10$ and $N_y=40$ cells. 
In the first $50\,\tau_\Lambda$, we let the system reach steady state under a fixed boundary condition and with anisotropic active cell stress with the required angle of the nematic field, $\Psi$. 
We then start the deformation of the box under pure shear with the required shear rate.
We propagate the system under pure shear until the values of the components of the mean cell elongation tensor have plateaued.
Then we let the system run for 10$\tau_\Lambda$ and calculate the time averaged value of the mean cell elongation.
This gives the steady-state values of the magnitude and angle of the mean elongation tensor.

\subsection{T1 transitions}
In this implementation of the vertex model, a full T1 transition is divided into 2 steps. 
First, two connected three-fold vertices that reach a distance $\mathcal{L}_{mn}$ lower than a certain threshold value, 
will merge to form a four-fold vertex. 
Next, a tentative split into two vertices is attempted in both possible topologies of the network. 
If in at least one of the topologies the forces acting on the vertices drives them apart, 
the four-fold vertex is replaced by two new three-fold vertices ~\cite{merkel2014cells, spencer2017vertex}. 
In case that cell neighbors have changed as compared to before the merger of the vertices, we call it a full T1 transition. 
Upon a T1 transition, which creates of a new bond at time $t_0$, the initial value of the bond tension $\Lambda_{mn}(t_0)$ is drawn from a normal distribution with mean $\bar{\Lambda}_{mn}$ and variance $\Delta\Lambda$.
After the new bond is created, the system is again relaxed to a force-balanced state.

\subsection{Constant cell-number ensemble}

In all of the vertex model simulations in this paper, we use a fixed cell number ensemble.
In rare cases however, bond tension fluctuations can drive the area of a cell below a critical area.
Below this critical area, the cell area imposed by the local minimum of this cell's work function is zero.
The cell would therefore shrink to have an area that is zero.
When the cell area reaches a value below a certain set threshold, the cell is extruded from the network and replaced by a vertex which has the same order as the neighbor number of the removed cell.
In order to keep the number of cells in the tissue constant, a randomly chosen cell divides.

In case a cell divides, a new bond is created running through the cell center.
The new bond makes an angle with the $x$-axis which is drawn from a uniform distribution between 0 and $\pi$. 
The two daughter cells have the same preferred area as the mother cell, simplifying the more realistic case of the continuous growth of the cell area. 
After each cell division, the configuration of the cell network is changed in order to minimize the work function. 

\section{Dual lattice triangulation and subcellular triangulation}

\label{sec_subcellular_triangulation}

In this appendix, we detail how the tissue shear decomposition, given by Eqs.~\eqref{eq_shear_decomp} and~\eqref{eq_shear_decomp_2} in the main text, differ in the \textit{dual lattice} and \textit{subcellular} triangulations.

\subsection{Contribution of a single T1 transition to tissue shear}

Here we describe how we obtained Fig.~\ref{fig_network_dynamics}, panels C and D, where we compare the shear decompositions of the dynamics of a T1 transition using a dual lattice triangulation and a subcellular triangulation.
The tissue is initialized with $10\times10$ regular hexagonal cells at $t=0$.
Panel C shows only the four cells connected to the bond involved in the T1 transition.
To model the T1 transition, we take two connected vertices in the cell network, and linearly decrease their distance in 50 steps until they lie on top of each other at $t=t_k$. 
After each step, we allow all vertices to move such that the network reaches a force-balanced state.
At time $t_k$ we change the topology of the cell network (T1 transition), and linearly increase the distance between the two new vertices until it reaches the rest length in the ground state at time $T$.

In panel D we show the shear decomposition of these dynamics using the dual lattice triangulation (top) and the subcellular triangulation (bottom), where all the quantities are averaged over the whole tissue.
Note that because the tissue is not deformed, the total shear is zero at all times.
Remarkably, in the subcellular triangulations, the shear by T1 transitions vanishes and the change in the mean cell elongation is compensated only by correlation effects.
This is explained below.

\subsection{Shear decomposition in the subcellular triangulation scheme}

Here we show that the mean elongation tensor $Q_{ij}$ is well defined for the subcellular triangulation and explain why the contribution to tissue shear stemming from T1 transitions $T_{ij}$ is zero.

To show that in the subcellular triangulation $Q_{ij}$ is finite at all times, let us consider a bond in the cellular network with length $L$ that undergoes a T1 transition.
During the T1 transition, the length $L$ gradually decreases to zero and a new bond with length zero is created in a new topology of the cellular network.
In the subcellular triangulation, the bond that undergoes a T1 transition is shared by two triangles.
The contribution to the area-weighted mean elongation of each of these two triangles, labeled $n$, is $A^n |\bm Q^n| \hat{Q}_{ij}$, with $A^n$ the triangle area, $|\bm Q^n|$ the magnitude of the triangle elongation and $\hat{Q}_{ij}$ the unit elongation tensor.
As $L$ decreases to zero, the area $A^n$ of the triangle scales as $A^n\sim L$.
Furthermore, taking the limit of small $L$ in the expression for the magnitude of the triangle elongation tensor, as defined in Eq. (A12) of Ref.~\cite{merkel2017triangles}, we find that $|\bm Q^n|\sim\log(L)$.
Because the area decreases faster than the elongation diverges as $L\to0$, we find
\begin{eqnarray}
 \lim_{L\to0} A^n |Q ^n|=0 \, .
 \label{Eq_limT1}
\end{eqnarray}

To conclude that $T_{ij}=0$ in the subcellular triangulation, notice that at the instantaneous event of a T1 transition, two triangles are removed from the mesh and two new triangles are created which all have an edge with a length zero. 
It follows from Eq.~\eqref{Eq_limT1} that these triangles do not contribute to the value of $Q_{ij}$, and thus that $Q_{ij}$ does not change at a T1 transition. Therefore, using the definition for $T_{ij}$ given in Eq.~(A44) in Ref.~\cite{merkel2017triangles}, it follows that T1 transitions do not contribute to tissue shear in the subcellular triangulation.

\section{Notation and definitions}
\label{sec_definitions}

        \subsection{Velocity gradient tensor}
        
The motion of cells in the tissue is described by the coarse-grained cell velocity field $v_j$ (or $V_j$ for the vertex model). Deformations of the network are proportional to gradients in this velocity field $\strainv{} = \partial_i v_j$, where $\strainv{}$ is the velocity gradient tensor. The trace of this tensor, $v_{kk}$ (summation over repeated Cartesian indices is implied), corresponds to local growth of the tissue. The traceless-symmetric part of the velocity gradient tensor, denoted $\tilde v_{ij}$, corresponds to anisotropic deformations, and its antisymmetric part, the vorticity tensor $\omega_{ij}$, characterizes local rotations. The velocity gradient tensor can thus be decomposed as:
\begin{equation}
 \strainv{} = \frac{1}{2} v_{kk} \delta_{ij} + \tilde v_{ij} + \omega_{ij},
 \label{eq_cont_velgrad_decomp}
\end{equation}
where in two dimensions we have $\omega_{ij}=-\omega\varepsilon_{ij}$ with $\varepsilon_{ij}$ the generator of counterclockwise rotation with $\varepsilon_{xy} = -1$, $\varepsilon_{yx}=1$ and $\varepsilon_{xx}=\varepsilon_{yy}=0$.

        \subsection{Stress tensor}

Similarly to the velocity gradient tensor, the tissue stress tensor $\sigma_{ij}$ can also be decomposed into
\begin{align}
    \sigma_{ij} = \frac{\sigma_{kk}}{d}\delta_{ij} + \tilde \sigma_{ij} \, .
\end{align}
where $d$ is the spatial dimension and $\tilde \sigma_{ij}$ is the shear stress. Note that in the absence of chiral terms the stress tensor is symmetric and we have therefore not included the  anti-symmetric contribution to the previous equation.

        \subsection{Corotational time derivative of tensors}

The corotational time derivative of a tensor $M_{ij}$ is defined as:
\begin{equation}
 \DDt{M_{ij}} = \ddt{M_{ij}} + \omega_{ik} M_{kj} + \omega_{jl} M_{il}.
 \label{eq_SI_hydr_model_corotational_derivative_simple_shear}
\end{equation}
where $\omega_{ij}$ is the vorticity tensor of the fluid.

        \subsection{Nematic tensors in two dimensions}
    \label{sec_nematic_tensor_decomposition}
    
In dimension two, a traceless symmetric tensor $M_{ij}$ (that we call nematic tensor) has two degrees of freedom and can be written in terms of its Cartesian coordinates as:
\begin{align}
    \bm{M} = 
    \begin{pmatrix}
    M_{xx} & M_{xy}  \\
    M_{xy} & -M_{xx}
    \end{pmatrix} \, ,
\end{align}
or it can equivalently be decomposed into a norm $M=|\bm{M}|$ and angle $\Theta$ as:
\begin{align}
    \bm{M} = M
    \begin{pmatrix}
    \cos(2\Theta) & \sin(2\Theta)  \\
    \sin(2\Theta) & -\cos(2\Theta) 
    \end{pmatrix} \, , \label{eq_definition_nematic_2d}
\end{align}
with $M=\sqrt{M_{xx}^2 + M_{xy}^2}$ and $\Theta=\frac{1}{2} \arctan (M_{xy},M_{xx})$, where the function $\arctan$ gives the arc tangent of $M_{xy}/M_{xx}$, taking into account in which quadrant the point $(M_{xy},\, M_{xx})$ lies. From Eq.~\eqref{eq_definition_nematic_2d}, one directly sees that $\bm M^2=M^2\mathbb{1}$ for a nematic tensor in two dimensions.

\section{Continuum theory of tissue dynamics}

Here we give an overview of the continuum models used in this paper. Similarly to the vertex model shear decomposition as given by Eq.~\eqref{eq_shear_decomp}, the tissue shear deformation rate $\tilde v_{ij}$ is decomposed in the continuum model as changes in cell shape $q_{ij}$ and cellular rearrangements $r_{ij}$ as:
\begin{align}
 \tilde v_{ij} = \DDt{q_{ij}} + r_{ij} \, ,
 \label{eq_cont_shear_decomp_SI}
\end{align}
which corresponds to Eq.~\eqref{eq_cont_shear_decomp} in the main text and that we have recalled here for convenience. The steady-state cell elongation depends crucially on the imposed boundary conditions, and we first discuss below the pure shear and simple shear boundary conditions as used in this paper. 

    \subsection{Boundary conditions in the continuum model}

        \subsubsection{Pure shear boundary condition}

Under pure shear, both the isotropic deformation and the global rotation rate vanish
$v_{kk}=0$, $\omega=0$, such that, assuming that the pure shear is applied in the direction $x$-axis, the global deformation-rate tensor reads:
\begin{eqnarray}
 v_{ij} = \tilde v_{ij} = 
 \begin{pmatrix}
  v_0 & 0 \\
  0 & -v_0
 \end{pmatrix}.
 \label{eq_BC_pure_shear}
\end{eqnarray}
Note that since the system is rotation free, the corotational derivative simplifies to a full time derivative $\DDtinline{q_{ij}}=\ddtinline{q_{ij}}$ in this case.

        \subsubsection{Simple shear boundary condition}
        
We consider here simple shear applied along the diagonal axis of the system ($xy$-axis). The deformation rate tensor and its decomposition reads:
\begin{eqnarray}
 v_{ij} = 
 \begin{pmatrix}
  0    & 0 \\
  2v_0 & 0
 \end{pmatrix} = \tilde v_{ij}  -\omega \varepsilon_{ij}  ,
\end{eqnarray}
and we identify $\tilde v_{xx} = 0,\, \tilde v_{xy} = v_0$ and the global rotation rate $\omega=-v_0$, see Eq.~\eqref{eq_cont_velgrad_decomp}. Under simple shear the system is not rotation-free and
the corotational derivative of the elongation tensor 
thus reads:
\begin{equation}
 \DDt{q_{ij}} = \ddt{q_{ij}} + \omega_{ik}q_{kj} + \omega_{jl}q_{il}.
 \label{eq_corotational}
\end{equation}
where $\omega_{ij}=-\omega \varepsilon_{ij}$.

    \subsection{Linear isotropic model}

\label{sec_hydr_model_linear_isotropic_shear}

We first consider a linear isotropic model for the dynamics of the tissue, where both the shear stress $\tilde \sigma_{ij}$ and tissue shear caused by topological rearrangements $r_{ij}$ are linear in the cell elongation tensor $q_{ij}$:
\begin{align}
    \tilde \sigma_{ij} = \shearmod{} \, q_{ij} \, , \quad  r_{ij} = k_1 \, q_{ij} \, ,
    \label{eq_linear_isotropic_SI}
\end{align}
which are Eqs.~\eqref{eq_shearModulus} and~\eqref{eq_linear_isotropic_rij} in the main text and that we recall here for convenience.

        \subsubsection{Dynamics under pure shear}

In the case of a \emph{pure shear} boundary condition, the time-dependent solution of the cell elongation $q_{ij}(t)$ can be obtained by solving the shear decomposition Eq.~\eqref{eq_cont_shear_decomp_SI} with the boundary condition~\eqref{eq_BC_pure_shear} and assuming linear constitutive equations given by Eq.~\eqref{eq_linear_isotropic_SI}. Assuming that the network is isotropic at $t=0$, we have the initial condition $q_{ij}(t=0)=0$ and we obtain:
\begin{equation}
    q_{xx}(t)=\frac{v_0}{k_1} \left(1 - e^{-k_1 t}\right), \quad q_{xy}(t)=0,
    \label{eq_SI_hydr_model_isotropic_linear_pure_shear_dyn}
\end{equation}
which converges to the steady-state cell elongation $q_{ij}^{\rm ss}$ for $t\to\infty$:
\begin{equation}
    q_{xx}^{\rm ss}=\frac{v_0}{k_1}, \quad q_{xy}^{\rm ss}=0 \, .
    \label{eq_SI_hydr_model_isotropic_linear_pure_shear_ss}
\end{equation}

        \subsubsection{Dynamics under simple shear}

Under \emph{simple shear}, the boundary condition is specified in Eq.~\eqref{eq_SI_hydr_model_corotational_derivative_simple_shear}. The time-dependent relaxation of the components of $q_{ij}(t)$ are obtained as
\begin{subequations}
\begin{align}
    q_{xx}(t)&=\frac{2v_0^2 - v_0 e^{-k_1 t} [v_0 \cos{(2v_0t)} + k_1 \sin{(2v_0t)}] }{k_1^2+4v_0^2},  \\
    q_{xy}(t)&=\frac{k_1v_0+ v_0 e^{-k_1 t} [2v_0\sin{(2v_0t)} - k_1 \cos{(2v_0t)}] }{k_1^2+4v_0^2}.
    \label{eq_SI_hydr_model_isotropic_linear_simple_shear_dyn}
\end{align}
\end{subequations}%
The steady-state solution reads:
\begin{equation}
    q_{xx}^{\rm ss} =\frac{2v_0^2}{k_1^2+4v_0^2}, \quad
    q_{xy}^{\rm ss} =\frac{k_1v_0}{k_1^2+4v_0^2}.
    \label{eq_SI_hydr_model_isotropic_linear_simple_shear_ss}
\end{equation}

    \subsection{Nonlinear isotropic model}
    \label{sec_hydr_model_non_linear_isotropic_shear}

In this section we now consider the nonlinear extension of the tissue constitutive equations, as introduced in Eq.~\eqref{eq_hydr_model_non_linear_isotropic} and that we recall here for convenience:
\begin{subequations}
\begin{align}
 r_{ij} &= k_1 \, q_{ij} + k_3 \mathrm{Tr}\left[ \bm q \cdot \bm q\right]q_{ij} \, ,  \label{eq_SI_hydr_model_isotropic_non_linear_rearrangements} \\
 \tilde \sigma_{ij} &= \shearmod{} \, q_{ij}\, .
\end{align}  \label{eq_SI_hydr_model_isotropic_non_linear}
\end{subequations}%
Note that a linear constitutive relation is still assumed for the shear stress. Below we discuss the steady-state solutions of Eq.~\eqref{eq_cont_shear_decomp_SI} under different boundary conditions and assuming the nonlinear constitutive equations as given by~Eq.~\eqref{eq_SI_hydr_model_isotropic_non_linear}.

    \subsubsection{Dynamics under pure shear}
    
In the case of a \emph{pure shear} boundary condition, the time-dependent solution of the cell elongation $q_{ij}(t)$ can be obtained by solving numerically the shear decomposition Eq.~\eqref{eq_cont_shear_decomp_SI} with the boundary condition~\eqref{eq_BC_pure_shear} and assuming the nonlinear constitutive equation given by Eq.~\eqref{eq_SI_hydr_model_isotropic_non_linear_rearrangements}. Using the numerical solutions of the resulting differential equation, we have fitted the data from the vertex model simulations of Fig.~\ref{fig_isotropic_pure_shear_triangulation_comparison}. The results of the fit are displayed in Fig.~\ref{fig_nonlinear_fit} and show an excellent agreement with the data, and should be compared with the similar fit obtained with the linear model (Fig.~\ref{fig_isotropic_pure_shear_hydr_model_fit}).

    \subsubsection{Steady-state solutions and linear stability analysis under pure shear}

We now focus on the steady-state solutions. Under a \emph{pure shear} deformation, the steady-state solutions obey $\tilde v_{ij}= r_{ij}$, and the steady-state elongation of a nonlinear tissue thus has a vanishing $xy$ component ($q_{xy}^{\rm ss} = 0$), while the $xx$ component is a root of the following polynomial:
\begin{align}
    f(q_{xx}) = q_{xx} \left(2 k_3 q_{xx}^2+k_1\right)- v_0 \, .
\end{align}
Since this polynomial is of third order in $q_{xx}$, it has three (possibly complex) roots. To discuss these roots and their stability, we focus on the case $v_0 >0$, since the roots of $f(q_{xx})$ remain unchanged under the transformation $v_0 \to - v_0$ and $q_{xx} \to - q_{xx}$.

Since $k_3>0$, the polynomial $f$ is monotonically increasing if $k_1>0$ and thus has a single real root which reads:
\begin{align}
    q_{xx}^{(1)} = \frac{\left(9 \sqrt{k_3} v_0+\sqrt{81 k_3 v_0^2+6 k_1^3}\right){}^{2/3}-\sqrt[3]{6} k_1}{6^{2/3} \sqrt{k_3} \sqrt[3]{9 \sqrt{k_3} v_0+\sqrt{81
   k_3 v_0^2+6 k_1^3}}} \, .
   \label{eq_real_root}
\end{align}

For $k_1<0$, $f$ has three real roots for $0\leq v_0<-k_1 \sqrt{-2k_1/3k_3}/3$ and a single real root for larger~$v_0$. The expression of the three real roots are lengthy and we give them in the limit of a small $v_0$ for simplicity. At first order in $v_0$, they read:
\begin{subequations}
\begin{align}
    q_{xx}^{(1)} &= \frac{v_0}{k_1} + \mathcal{O}(v_0^2) \, , \\
    q_{xx}^{(2,3)} &= \frac{v_0}{- 2 k_1} \pm \sqrt{\frac{-k_1}{2k_3}} + \mathcal{O}(v_0^2) \, . \label{eq_stable_root}
\end{align}
\end{subequations}%
We now study the stability of these different solutions by considering an infinitesimal perturbation around the steady-state solution:
\begin{subequations}
\begin{align}
    q_{xx}(t) &= q_{xx}^{\rm ss} +\delta q_{xx}(t) \, ,\\
    q_{xy}(t) &= q_{xy}^{\rm ss}  +\delta q_{xy}(t)  \, ,
\end{align} \label{eq_linear_perturbation}%
\end{subequations}%
where $q_{xx}^{\rm ss}=q_{xx}^{(1,2,3)}$ and $q_{xy}^{\rm ss}=0$.
Injecting the expression~\eqref{eq_linear_perturbation} in the shear decomposition Eq.~\eqref{eq_cont_shear_decomp_SI} and expanding at first order in the perturbation, we obtain:
\begin{align}
    \partial_t  
    \begin{pmatrix}
    \delta q_{xx} \\ \delta q_{xy}
    \end{pmatrix} = S \, . 
    \begin{pmatrix}
    \delta q_{xx} \\ \delta q_{xy}
    \end{pmatrix}\, ,
\end{align}
where we have defined the stability matrix:
\begin{align}
    S = 
    \begin{pmatrix}
    - k_1 - 6 k_3 \left( q_{xx}^{\rm ss} \right)^2  & 0\\
    0 & -k_1 - 2 k_3 \left( q_{xx}^{\rm ss} \right)^2
    \end{pmatrix} \, ,
\end{align}
whose eigenvalues must both be negative in order to have a stable solution. 

The unique real root in the case $k_1>0$ given by Eq.~\eqref{eq_real_root} is therefore always stable since both eigenvalues are negative. In the case $k_1<0$ and for small imposed shear rate $v_0$, we deduce from the eigenvalues of the stability matrix that $q_{xx}^{(2)}$, given in Eq.~\eqref{eq_stable_root}, is the only stable steady-state solution.

As a result of this analysis and for vanishing external shear ($v_0=0$), we observe that our mean-field nonlinear tissue model undergo a supercritical bifurcation as $k_1$ goes from positive to negative values. The tissue is spontaneously elongated with $q_{xx} = \sqrt{-k_1/2k_3}$ as soon as $k_1<0$, even for vanishing external shear.

Importantly, since there exist a linear relationship between elongation and stress: $\tilde\sigma_{ij} = \shearmod{} \, q_{ij}$, the spontaneous elongation discussed above corresponds to a yield stress:
\begin{align}
    \sigma^{\rm y}_{xx} = \shearmod{} \sqrt{\frac{-k_1}{2k_3}} \, . 
\end{align}
At positive values of the shear rate $v_0$, the continuum model also predicts a shear thinning behavior as the slope of the shear stress as a function of the shear rate is a decreasing function of the shear rate. This is made explicit by looking at large shear rate, for which we find that the steady-state stress behaves as:
\begin{align}
    \tilde\sigma_{xx} \underset{v_0\to \infty}{\sim} (2 k_3)^{-1/3} \, v_0^{1/3}  \, ,
\end{align}
indicating a shear thinning at large shear rate..

    \subsubsection{Steady-state solutions under simple shear}

\begin{figure}[t]
 \centering
 \includegraphics[width=0.7\textwidth]{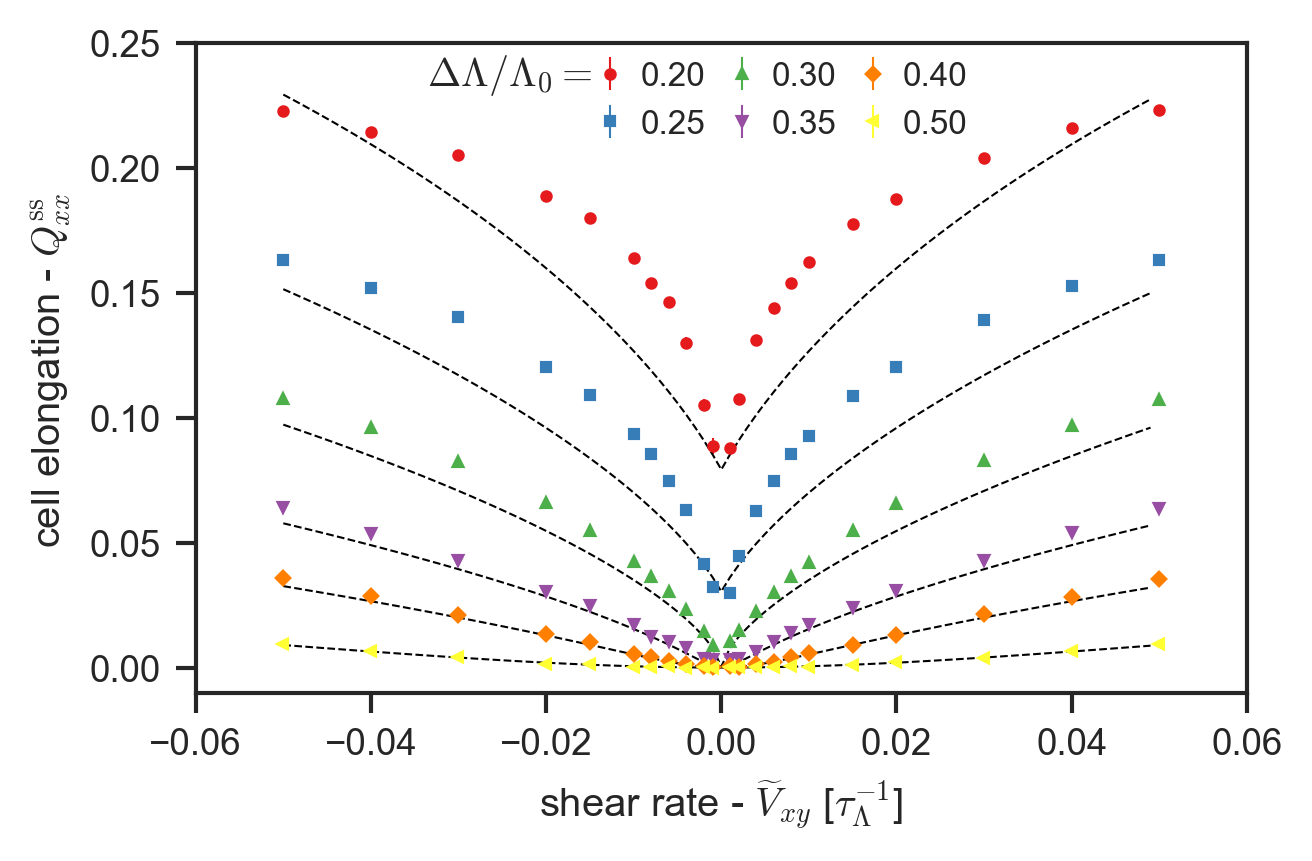}
 \caption{Steady-state value of the $xx$-component of the mean cell elongation tensor against the imposed simple shear rate, $\widetilde{V}_{xy}$. Different colors correspond to different magnitudes of the line tension fluctuations, $\Delta\Lambda$, given in the legend. The $xy$-component of the cell elongation tensor is given in Fig.~\ref{fig_isotropic_simple_shear_vary_bond_tensions}. Dashed black lines show fits of the nonlinear continuum model in Eq.~\eqref{eq_hydr_model_non_linear_isotropic}. Values obtained from the fits of the parameters $k_1$ and $k_2$ are given in Table~\ref{tab_hydr_model_isotropic_shear_mod}.}
 \label{fig_SI_isotropic_simple_shear_vary_bond_tensions}
\end{figure}

For a tissue with an imposed \emph{simple shear} deformation, there are also in general three steady-state solutions of Eqs.~\eqref{eq_cont_shear_decomp_SI} and~\eqref{eq_SI_hydr_model_isotropic_non_linear} for $q_{ij}$. Note that under \emph{simple shear}, the corotational derivative appearing in Eq.~\eqref{eq_cont_shear_decomp_SI}  is nontrivial and is computed using Eq.~\eqref{eq_corotational}.
The stability analysis of these solutions can be carried out similarly to the pure shear case, either numerically or analytically in the limit of a small shear rate, and we present the latter below, assuming that $v_0>0$. The outcome of this linear stability analysis is the existence of a single stable steady-state solution, which reads:
\begin{subequations}
\begin{align}
    q_{xx}^{\rm ss} &= \frac{-12 k_3^2 v_0^2 + D_3^2}{3 k_3^2 D_0^{1/3}} \, , \\
    q_{xy}^{\rm ss} &= \frac{48 v_0^4 k_3^4 + \tfrac{1}{3} D_1^{1/2} D_3 + v_0^2 k_3^2 \left( (16 k_1 k_3+9 k_3^2) D_0^{1/3} - 20 k_1^2 k_3^2 - 9 k_1 k_3^3 - 8 D_0^{2/3} \right)}{6 v_0 k_3^3 D_0^{2/3}} \, , 
\end{align} \label{non_linear_hydr_model_isotropic_simple_shear_solution}
where we have introduced the following parameters:
\begin{align}
\begin{split}
 D_0 &= \sqrt{D_1} + D_2 \, ,  \\
 D_1 &= 27 \left( v_0^2 k_3^6 \left[ 4 (4 v_0^2+k_1^2)^2 + 2 k_1 k_3 (36 v_0^2 + k_1^2) + 27 v_0^2 k_3^2 \right] \right) \, ,  \\
 D_2 &= k_3^3 \left( k_1^3 + 9 v_0^2 (4 k_1 + 3 k_3) \right) \, ,  \\
 D_3 &= D_0^{1/3} - k_1 k_3 \, .    
\end{split}
\end{align}
\end{subequations}%
The solution given by Eq.~\eqref{non_linear_hydr_model_isotropic_simple_shear_solution} has been used to obtain the fits given in Figs.~\ref{fig_isotropic_simple_shear_vary_bond_tensions} and~\ref{fig_SI_isotropic_simple_shear_vary_bond_tensions}, with the corresponding fitted parameters given in Table~\ref{tab_hydr_model_isotropic_shear_mod}. Note that this solution exists and is stable if $k_1\geq0$, or if $k_1<0$ and $k_3\geq-2k_1$. In the case where $k_1<0$ and $k_3<-2k_1$, there is no linearly stable solution.

\begin{table}[t]
  \begin{tabular}{ |c||r|r| }
    \hline
    & \multicolumn{2}{|c|}{Parameter values from fit} \\
    \hline
    $\Delta\Lambda/\Lambda_0$ & $k_1\, [\tau_\Lambda^{-1}]$ & $k_3\, [\tau_\Lambda^{-1}]$ \\
    \hline 
    0.20 & -0.06 &   0.72 \\
    0.25 & -0.04 &   1.25 \\
    0.30 &  0.00 &   2.10 \\
    0.35 &  0.05 &   3.90 \\
    0.40 &  0.16 &   6.75  \\
    0.50 &  0.61 &  14.06  \\
    \hline
  \end{tabular}
  \caption{Values of obtained by fitting the nonlinear continuum model of an isotropic tissue with constitutive equation~\eqref{eq_SI_hydr_model_isotropic_non_linear} to the steady-state values of the mean cell elongation obtained from simulations of a cellular network. Fitted values were obtained for different magnitudes $\Delta\Lambda$ of the bond tension fluctuations.}
  \label{tab_hydr_model_isotropic_shear_mod}
\end{table}

We now give details on the linear stability analysis in the limit of small shear rate. When $k_1>0$, only one of the three solutions is real and is obtained by expanding Eq.~\eqref{non_linear_hydr_model_isotropic_simple_shear_solution} in series of the shear rate $v_0$. It reads:
\begin{subequations}
\begin{align}
    q_{xx}^{\rm ss} &= 0 + \mathcal{O}(v_0^2) \, , \\
    q_{xy}^{\rm ss} &= \frac{v_0}{k_1} + \mathcal{O}(v_0^2) \, .
\end{align}
\end{subequations}%
and the eigenvalues of the stability matrix are $\{-k_1+ \mathcal{O}(v_0^2),-k_1+ \mathcal{O}(v_0^2) \}$ indicating that this solution is linearly stable.

When $k_1<0$, there is a unique stable solution which reads in the limit of small shear rate:
\begin{subequations}
\begin{align}
    q_{xx}^{\rm ss} &= -\frac{k_1}{k_3} + v_0 \frac{ \sqrt{2(2 k_1+k_3)}}{k_3\sqrt{-k_1}} + \mathcal{O}(v_0^2) \, , \\
    q_{xy}^{\rm ss} &= \frac{\sqrt{-k_1(2 k_1+k_3)}}{\sqrt{2}\, k_3}-\left(\frac{2}{k_3}+\frac{1}{2 k_1}\right) v_0+ \mathcal{O}(v_0^2) \, .
\end{align}
\end{subequations}%
Note that this solution is real provided $k_3\geq-2k_1$. There is no stable solution for $k_1<0$ and $k_3<-2k_1$. The eigenvalues of the stability matrix corresponding to this solution are:
\begin{align}
2 k_1 + v_0\frac{3 \sqrt{-k_1 (4 k_1+2 k_3)}}{k_1} + \mathcal{O}(v_0^2), \, v_0\frac{\sqrt{-k_1 (4 k_1+2k_3)}}{k_1} +  \mathcal{O}(v_0^2) \, ,
\end{align}
and they are both negative when  $k_1<0$ and $k_3\geq-2k_1$, showing that the solution is indeed stable.

    \subsection{Continuum theory with active anisotropic stress}
    
    \label{sec_SI_anisotropic}

In the continuum model with active anisotropic stress, 
we introduce a nematic tensor $p_{ij}$, parameterized by the angle $\psi$ with respect to the $x$-axis, that sets the preferred axis in the tissue along which the active stress is exerted.
In all models described below, we consider the case of a constant cell polarity with unit magnitude, such that the components of $p_{ij}(\psi)$ are $p_{xx} =- p_{yy} =\cos{2\psi}$ and $p_{xy}  = p_{yx}  =\sin{2\psi}$.

\subsubsection{Linear theory with active anisotropic stress}

\alabel{SI_hydr_model_linear_anisotropic_stress}

The constitutive equations for an anisotropic cell network read, at first order in the mean cell elongation $q_{ij}$ and in the cell nematic field $p_{ij}$:
\begin{subequations}
\begin{align}
    r_{ij} &= k_1 \, q_{ij}  + \lambda \,  p_{ij} \, ,  \\
    \tilde\sigma_{ij} &= \shearmod \, q_{ij}+ \zeta \, p_{ij} \, ,
\end{align}  \label{eq_SI_hydr_model_anisotropic_stress_linear}%
\end{subequations}%
where $\lambda$ and $\zeta$ are the magnitude of the active rearrangements and active shear stresses in the tissue, respectively.
For a tissue under an imposed \emph{pure shear} deformation, the solution of Eq.~\eqref{eq_cont_shear_decomp_SI} with the boundary condition~\eqref{eq_BC_pure_shear} and the constitutive equations~\eqref{eq_SI_hydr_model_anisotropic_stress_linear} reads: 
\begin{align}
    q_{ij}(t) = \frac{1}{k_1}\left(1-e^{-k_1 t}\right)\tilde v_{ij} - \frac{\lambda_0}{k_1} \left(1-e^{-k_1 t} \right) p_{ij} \, .
 \label{eq_SI_hydr_model_anisotropic_stress_linear_pure_shear}
\end{align}
with the initial condition $q_{ij}(t=0) = 0$.
In steady state, Eq.~\eqref{eq_SI_hydr_model_anisotropic_stress_linear_pure_shear} reduces to
\begin{equation}
    q_{ij}^{\rm ss} = \frac{1}{k_1}\left(\tilde v_{ij} - \lambda_0 p_{ij} \right).
\end{equation}

The case of a cell network under \emph{simple shear} is treated similarly, using the boundary condition~\eqref{eq_SI_hydr_model_corotational_derivative_simple_shear}.
We obtain the following steady-state solutions for the cell elongation:
\begin{equation}
    q_{xx}^{\rm ss} = \frac{2v_0\left( v_0 - 2\lambda_0\sin{2\psi}\right) - k_1 \lambda_0\cos{2\psi}}{k_1^2 + 4  v_0^2} \, , \quad 
    q_{xy}^{\rm ss}= \frac{k_1\left(v_0 - \lambda_0 \sin{2\psi} \right) + 2v_0\lambda_0\cos{2\psi}}{k_1^2 + 4v_0^2} \, .
\end{equation}

\subsubsection{Nonlinear theory with active anisotropic stress}

As discussed in the main text, one can go beyond the linear model in Eq.~\eqref{eq_SI_hydr_model_anisotropic_stress_linear} 
by including a nonlinear dependence of $r_{ij}$ on the mean cell elongation $q_{ij}$. To this end, we include all contractions between $q_{ij}$ and $p_{ij}$ that are allowed by the symmetries in the system and that yield a symmetric traceless tensor. We find, at second order in $|\bm q|$, that the most general expression for $r_{ij}$ for a two-dimensional system reads:
\begin{align}
\begin{split}
    r_{ij} &= 
    \left(k_1 + \frac{k_2}{2}\mathrm{Tr}[\bm p\cdot \bm q] \right) q_{ij} 
    + \left( \gamma_0 + \frac{\gamma_1}{2}\mathrm{Tr}[\bm p\cdot \bm q] \right) p_{ik} q_{kl} p_{lj}
    +  \delta_0 \, p_{ik}q_{kl}p_{lm}q_{mn}p_{nj} \\
    &+ \left( \lambda_0 + \frac{\lambda_1}{2}\mathrm{Tr}[\bm p\cdot \bm q] 
    + \frac{\lambda_2}{2} \mathrm{Tr}[\bm q\cdot \bm q] 
    + \frac{\lambda_3}{2}\mathrm{Tr}[\bm p\cdot \bm q \cdot \bm p\cdot \bm q] 
    + \frac{\lambda_5}{4}\mathrm{Tr}[\bm p\cdot \bm q]^2 \right) p_{ij}
    + \mathcal{O}(|\bm q|^3 )\, ,
\end{split}\label{eq_SI_hydr_model_anisotropic_stress_non_linear2}
\end{align}
which has 10 parameters.

In order to find a closed solution for $q_{ij}$, we first parameterize the mean elongation tensor in terms of its norm $q=|\bm q|$ and its angle $\phi$ (see App.~\ref{sec_nematic_tensor_decomposition}). We then expand the dependence on the angle of cell elongation 
$\phi$ close to the angle of the cell polarity tensor $\psi$ as: $\delta\phi = \phi-\psi$, such that $q_{ij}(q,\phi) = q_{ij}(q, \psi+\delta\phi)$. Substituting this expression into Eq.~\eqref{eq_SI_hydr_model_anisotropic_stress_non_linear2} and expanding in series of $\delta\phi$, we obtain for the components of $r_{ij}$ at first order:
\begin{equation}
    r_{ij}^{(1)} = \left[ g(q)\,\delta_{ik} + 2\delta\phi \,h(q) \, \varepsilon_{ik} \right] p_{kj} \, ,
    \label{eq_SI_hydr_model_anisotropic_stress_linearized2}
\end{equation}
where $g(q)=(\beta_2/2 q^2 + \beta_1 q + \lambda_0)$ and $h(q)=(\beta_4 q^2+\beta_3 q+\lambda_0)$. Here we have introduced the parameters $\beta_i$, which, expressed in the original parameters of the model, read:
\begin{equation}
 \beta_1 = k_1 + \gamma_0 + \lambda_1, \quad \beta_2 = 2(k_2+\gamma_1+\delta_0+\lambda_2 + \lambda_3 + \lambda_5), \quad \beta_3 = k_1-\gamma_0, \quad \beta_4 = k_2-\gamma_1-2\delta_0 \, .
 \label{eq_SI_hydr_non_linear_anisotropic_parameters}
\end{equation}

Under an imposed \emph{pure shear} deformation, the steady-state shear decomposition reduces to $\tilde v_{ij} = r_{ij}$, which we can solve to obtain the expressions for the steady-state magnitude of cell elongation $q^{\mathrm{ss}}$, and the deviation of the elongation direction from the polarity axis, $\delta\phi^{\mathrm{ss}}$:
\begin{subequations}
\begin{align}
    q^{\mathrm{ss}}(v_0, \psi) &= \frac{1}{\beta_2} \left( -\beta_1 + \sqrt{\beta_1^2 + 2\beta_2[v_0 \cos(2\psi) - \lambda_0]} \right) \, , \\
 \delta\phi^{\mathrm{ss}}(v_0,\psi) &= \frac{v_0 \sin(2\psi)}{2q^{\mathrm{ss}}(\beta_3 + \beta_4 q^{\mathrm{ss}})}\, .
\end{align}  \label{eq_SI_hydr_model_anisotropic_stress_linearized2_pure_shear}
\end{subequations}%
We can fit these expressions for $q$ and $\delta\psi$ to the steady state values of the mean cell elongation in the vertex model for different angles of cell polarity $\psi$ and at different applied shear rates $v_0$ to find the values of the 5 parameters $\beta_{1,2,3,4}$ and $\lambda_0$. These fitted values are displayed in Table~\ref{tab_SI_hydr_model_anisotropic_stress_fitted_values}.

To obtain the sensitivity of the magnitude of cell elongation to changes in the applied shear rate, we take the derivative of the magnitude of steady state cell elongation, see Eq.~\eqref{eq_SI_hydr_model_anisotropic_stress_linearized2_pure_shear}, 
with respect to $\tilde v_{xx}$, we get: 
\begin{equation}
 \frac{\partial q^{\mathrm{ss}}}{\partial \tilde v_{xx}} = \frac{\cos(2\psi)}{\sqrt{\beta_1^2 + 2\beta_2 [\tilde v_{xx}\cos(2\psi) -\lambda_0]}}\, .
\end{equation}

        \subsubsection{Shear stress in the presence of tissue anisotropy}
        
        \label{sec_SI_anisotropic_stress}
        
\begin{figure}[t]
 \centering
 \includegraphics[width=0.95\textwidth]{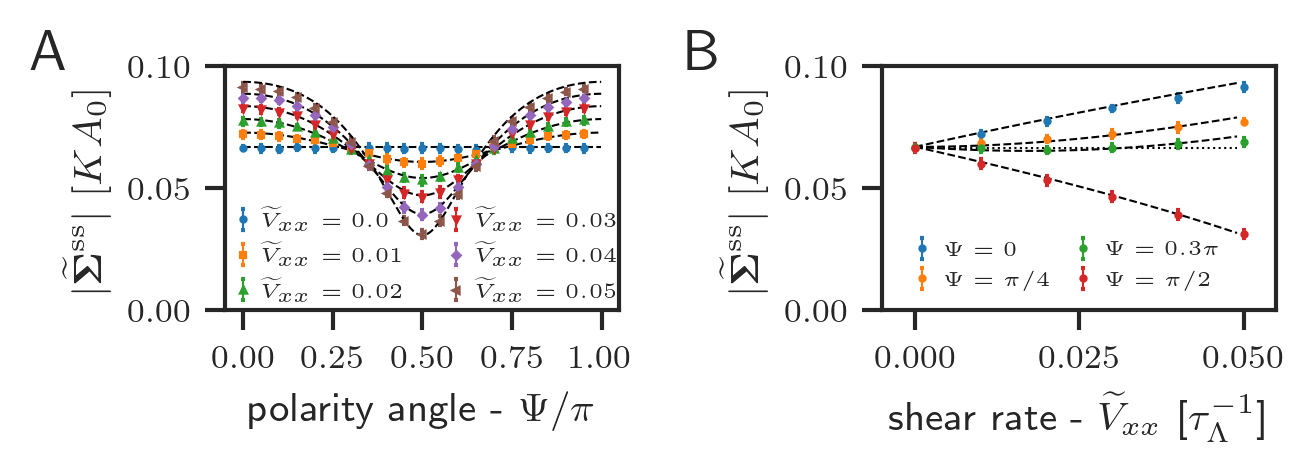}
 \caption{
 Magnitude of the steady state tissue shear stress tensor, 
 $|\widetilde{\bm{\Sigma}}^{\rm ss}|$, against different angles of the polarity tensor $\Psi$ (panel A) or different values of the applied pure shear rate $\widetilde{V}_{xx}$ (panel B).
 In both panels, circle markers indicate the magnitude of the shear stress tensor obtained from vertex model simulations and dashed black lines analytic predictions from the continuum model in Eq.~\eqref{eq_SI_hydr_model_anisotropic_stress_non_linear1_stress_mag}.
 In panel A, different colors indicate the rate of imposed pure shear, $\widetilde{V}_{xx}$, given in the legend, and in panel B differed colors indicate the angle of polarity, $\Psi$, given in the legend.}
 \label{fig_SI_hydr_model_anisotropic_stress_stress_tensor}
\end{figure}

In Fig.~\ref{fig_SI_hydr_model_anisotropic_stress_stress_tensor} we show the magnitude of the tissue shear stress tensor obtained from vertex model simulations, 
$|\tilde{\bm{\Sigma}}^{\rm ss}|=\tfrac{1}{2}\mathrm{Tr}[\widetilde{\bm{\Sigma}}^T\cdot\widetilde{\bm{\Sigma}}]$, at different pure shear rates $\tilde V_{xx}$, and different angles of the polarity field, $\Psi$.
A correct continuum model for the tissue shear stress in the presence of an anisotropic active stress at the cellular level requires the modification of its constitutive equation as compared to the isotropic case.
At linear order in cell elongation, we first write
\begin{align}
    \tilde \sigma_{ij} = \mu_\mathrm{s} q_{ij} + \xi_0 p_{ij} \, ,
    \label{eq_SI_hydr_model_anisotropic_stress_linear_stress}
\end{align}
where $p_{ij}$ is the nematic director field of the active stress and $\xi_0$ the magnitude of the active part of the shear stress. 
As we argue below, this form is missing couplings of the same order in $q$ and is not sufficient to capture the properties of the vertex model. The correct expression (at first order in $q$) will be given in Eq.~\eqref{eq_SI_hydr_model_anisotropic_stress_non_linear1_stress}.

In the continuum model and in the case of a linear constitutive equation for the stress as given in Eq.~\eqref{eq_SI_hydr_model_anisotropic_stress_linear_stress}, we find:
\begin{align}
 |\tilde{\sigma}|^2 = \mu_\mathrm{s}^2 q^2 + \xi_0^2 + 2 \mu_\mathrm{s} \xi_0 q \cos(2\delta\phi) \, .
 \label{eq_SI_hydr_model_anisotropic_stress_stress_magnitude}
\end{align}
Here, the dependence of the tissue shear stress on the shear rate and polarity angle is through the mean cell elongation magnitude $q$, and the change in elongation direction, $\delta\phi$.
Panel~B of Fig.~\ref{fig_SI_hydr_model_anisotropic_stress_stress_tensor} shows a clear dependence of the magnitude of the shear stress on the applied pure shear rate at $\Psi=\pi/4$ (orange points).
However, it is clear from Fig.~\ref{fig_active_stress_polarity_dependence_ss_fit}, panel A, that the magnitude of mean cell elongation $|\bm{Q}^{\rm ss}|$ does not depend on the applied shear rate at $\Psi=\pi/4$.
Therefore, the dependence on the shear rate at $\psi=\pi/4$ in Eq.~\eqref{eq_SI_hydr_model_anisotropic_stress_stress_magnitude} must come from the change in the angle of the mean cell elongation tensor $\delta\phi$.
However, the dependence on $v_0$ in Eq.~\eqref{eq_SI_hydr_model_anisotropic_stress_stress_magnitude} is only quadratic to lowest order (see Eq.~\eqref{eq_SI_hydr_model_anisotropic_stress_linearized2_pure_shear}).
As a result, we found that fitting Eq.~\eqref{eq_SI_hydr_model_anisotropic_stress_stress_magnitude} to the data in Fig.~\ref{fig_SI_hydr_model_anisotropic_stress_stress_tensor} is not satisfactory, as it would predict a negligible dependence of the shear stress on the applied shear rate at $\psi=\pi/4$.

In order to obtain an accurate description of the shear stress, we supplement the constitutive equation~\eqref{eq_SI_hydr_model_anisotropic_stress_linear_stress} for the shear stress with terms of the same order in the magnitude of the cell elongation tensor but that involve more contractions with the nematic tensor $p_{ij}$.
The correct constitutive equation at first order in $q$ reads:
\begin{equation}
    \tilde \sigma_{ij}  = \mu_{\mathrm{s}} q_{ij} + \left( \xi_0 + \xi_1 \mathrm{Tr}[\bm p \cdot\bm q] \right) p_{ij} + \chi_0 p_{ik} q_{kl}p_{lj} \, ,
\label{eq_SI_hydr_model_anisotropic_stress_non_linear1_stress}
\end{equation}
from which follows the computation of the shear stress magnitude:
\begin{align}
\begin{split}
 |\tilde{\sigma}|^2 &= \xi_0^2 + \left(\mu_\mathrm{s}^2 + \mu_s\xi_1 + \tfrac{1}{2}\xi_1^2 + \xi_1\chi_0 + \chi_0^2 \right)q^2 
 + 2 \xi_0 (\mu_\mathrm{s} + \xi_1 + \chi_0) q \cos(2\delta\phi) \\
 & + \frac{1}{2} q^2 (2\mu_s + \xi_1 )(\xi_1 + 2\chi_0) \cos(4\delta\phi) \, .
\end{split} \label{eq_SI_hydr_model_anisotropic_stress_non_linear1_stress_mag}
\end{align}
We can now fit Eq.~\eqref{eq_SI_hydr_model_anisotropic_stress_non_linear1_stress_mag} to the vertex simulations and we plot them as dashed line in Fig.~\ref{fig_SI_hydr_model_anisotropic_stress_stress_tensor}. 
The values obtained from the fits are given in Table~\ref{tab_SI_hydr_model_anisotropic_stress_fitted_values}.

\begin{table}[t]
  \begin{tabular}{|l|l|l|l|l|l|l|l|l| }
    \hline
    \multicolumn{9}{|c|}{Parameter values obtained from fit} \\
    \hline
      $\beta_1\, [\tau_\Lambda^{-1}]$ & $\beta_2\, [\tau_\Lambda^{-1}]$ & $\beta_3\, [\tau_\Lambda^{-1}]$ & $\beta_4\, [\tau_\Lambda^{-1}]$ &  $\lambda_0\, [\tau_\Lambda^{-1}]$ & $\shearmod{}\,[K A_0]$ & $\xi_0\, [K A_0]$ & $\xi_1\, [K A_0]$ &  $\chi_0\, [K A_0]$ \\
    \hline 
    -0.20 & 2.82 & 0.26 & 1.60 & -0.08 & 0.44 & -0.07 & -1.19 & 1.17 \\
    \hline
  \end{tabular}
  \caption{Values of the parameters in the continuum model for tissue shear flow and shear stress, presented in Eqs.~\eqref{eq_SI_hydr_model_anisotropic_stress_linearized2_pure_shear} and~\eqref{eq_SI_hydr_model_anisotropic_stress_non_linear1_stress}, obtained from fits to the steady-state cell elongation and shear stress values from simulations of cellular network, presented in Fig.~\ref{fig_SI_hydr_model_anisotropic_stress_stress_tensor}. The shear modulus \shearmod{} was obtained from a fit to the isotropic model, see Fig.~\ref{fig_isotropic_simple_shear_vary_bond_tensions}.}
  \label{tab_SI_hydr_model_anisotropic_stress_fitted_values}
\end{table}

\bibliographystyle{apsrev4-1}
\bibliography{biblio.bib,biblio_extra.bib}

\end{document}